\documentclass[aps,prd,twocolumn,superscriptaddress,nofootinbib]{revtex4-2}

\usepackage{graphicx}
\usepackage{dcolumn}
\makeatletter
\def\switch@array{}
\makeatother
\usepackage{bm}
\usepackage{amsmath}

\usepackage{amsfonts}
\usepackage{amssymb}
\usepackage[usenames,dvipsnames,svgnames]{xcolor}
\usepackage{hyperref}
\hypersetup{
    pdfnewwindow=true,
    colorlinks=true,
    allcolors=[RGB]{31 119 180}
}
\usepackage[capitalise]{cleveref}
\usepackage{placeins} 
\usepackage{nameref}
\usepackage{xpatch}
\usepackage{orcidlink}

\DeclareMathOperator{\re}{Re}
\DeclareMathOperator{\im}{Im}
\newcommand{\diff}{\ensuremath{\mathrm{d}}}
\newcommand\bsub{\begin{subequations}}
\newcommand\esub{\end{subequations}}

\makeatletter
\xpatchcmd{\@ssect@ltx}{\@xsect}{\protected@edef\@currentlabelname{#8}\@xsect}{}{}
\xpatchcmd{\@sect@ltx}{\@xsect}{\protected@edef\@currentlabelname{#8}\@xsect}{}{}
\makeatother
\graphicspath{{figures/}}

\usepackage{orcidlink}

\newcommand{\ub}{Departament de F\'isica Qu\`antica i Astrof\'isica and Institut de Ci\`encies del Cosmos, Universitat de Barcelona,  E-08028 Barcelona, Spain}

\newcommand{\glasgow}{School of Physics and Astronomy, University of Glasgow, Glasgow, G12 8QQ, UK}

\begin{document}

\title{ Elliptical Polarization in Partial Wave Analysis of Two Spinless Meson Photoproduction}


\author{D.~I.~\surname{Glazier}\orcidlink{0000-0002-8929-6332}}
\affiliation{\glasgow}

\author{V.~\surname{Mathieu}\orcidlink{0000-0003-4955-3311}}
\affiliation{\ub}

\begin{abstract}

Mathematical ambiguities in partial-wave analysis present a significant challenge to the extraction of resonance properties in hadronic reactions. Recent work has shown that while linear photon polarization can resolve continuous ambiguities in the photoproduction of two pseudoscalar mesons, a final complex conjugate ambiguity remains. In this work, we extend the partial-wave formalism to include circular and elliptical photon polarization. We demonstrate that the additional constraints provided by circularly polarized observables, which are sensitive to the imaginary parts of bilinear amplitude products, are sufficient to remove remaining mathematical ambiguities, yielding an improved partial-wave solution. Furthermore, we show that the resulting overconstrained system allows for a novel application: using the reaction dynamics themselves as a polarimeter. Using recent high-statistics data on $\rho(770)$ photoproduction from the GlueX experiment, we illustrate the viability of this technique in determining the degrees of beam polarization from the data. These results will benefit the next generation of photoproduction experiments at facilities such as Jefferson Lab and the future Electron-Ion Collider.

\end{abstract}
\maketitle
\section{Introduction}
\label{sec:intro}
In the pursuit of understanding the spectrum of hadrons, partial-wave analysis (PWA) stands as a critical tool. The main goal of experiments in this field is to extract the partial-wave amplitude content in particular final states. Subsequent analysis of the mass dependence of these amplitudes allows for the extraction of resonant poles, thereby providing access to the contributing parts of the hadronic spectrum. The formalism for such analyzes, particularly for reactions involving particle spin, rests on foundational works. The helicity formalism introduced by Jacob and Wick~\cite{Jacob:1959at} provides the standard language for describing these scattering processes, while the work of Schilling, Seyboth, and Wolf~\cite{Schilling:1970vj} established the essential framework for analyzing vector meson production with polarized photons in terms of Spin-Density Matrix Elements (SDMEs).

A well-known challenge in PWA is the potential for mathematical ambiguities, where multiple sets of partial-wave amplitudes can describe the experimental observables equally well. Barrelet \cite{Barrelet:1971pw} formulated an approach to study this based on the zeros of the scattering amplitude. There were precursors to this work on the extraction of amplitudes in pion-nucleon scattering by Gersten~\cite{GERSTEN1969537} and the extension to single meson photoproduction by Omelaenko~\cite{omelaenko1981sov}, demonstrating how limited sets of observables can constrain amplitudes up to discrete ambiguities.
More recently extensive studies, notably by S. U. Chung~\cite{Chung:1997qd}, have explored these ambiguities in reactions initiated by spinless beams, framing the problem in terms of Barrelet zeros and introducing the powerful concept of the reflectivity basis to separate amplitudes by the naturality of the t-channel exchange particle~\cite{Chung:1975qj}.

A study of truncated partial wave analysis of two spinless meson photoproduction focused on the complete experiment of polarization observable required for amplitude analysis of s-channel nucleon resonance production \cite{Fix:2022tfc}. It highlights the role of parity conservation and complex conjugate ambiguities.
Recent publications have extended these investigations to the domain of mesonic resonance photoproduction, demonstrating how polarized photon beams can constrain partial-wave amplitudes of intermediate meson resonances decaying into two spinless particles~\cite{PhysRevD.108.076001,kjcp-h8b9}. These studies have focused on linearly polarized photons, largely because of the recent operation of the GlueX experiment. It was shown that the use of linear polarization in the t-channel production of two spinless mesons provides two significant benefits compared to unpolarized production: amplitudes may be further split by their reflectivity, and ambiguities in the extracted partial waves are reduced significantly, up to a complex conjugate, effectively constraining the partial wave content up to averaging over the proton spin.\footnote{Denoting by $[\ell]^{(\epsilon)}_{m;k = 0}$ and $[\ell]^{(\epsilon)}_{m;k = 1}$ the nucleon non-flip and nucleon-flip amplitudes of the wave $\ell$, projection $m$ and reflectivity $\epsilon$, without any information about the polarization of the initial or final proton, the observables are all proportional to $\left(|[\ell]^{(\epsilon)}_{m;0}|^2 + |[\ell]^{(\epsilon)}_{m;1}|^2 \right)^{1/2} \equiv [\ell]^{(\epsilon)}_{m}$ . We are calling this combination the average over the proton spin.}

However, future GlueX running~\cite{hurck2024proposalpac52measurement} and other experiments such as CLAS12~\cite{ACKER2020163475} and the proposed EIC~\cite{ABDULKHALEK2022122447} will also have effective circular polarization from quasi-real photoproduction. Here we consider the implications of a circularly polarized photon beam on the partial-wave analysis of these meson decays. While the addition of linear polarization already provides powerful constraints, it does not mean that information from circular polarization is not also useful; an overconstrained system can provide sensitive tests of the underlying model or be used to determine external parameters such as the degree of beam polarization itself.


\section{Partial Waves and Spherical Harmonic Moments}
\subsection{Partial Wave Formalism}
\label{sec:formalism}

\begin{figure}[tbp]
\begin{center}
\includegraphics[width=0.9\linewidth]{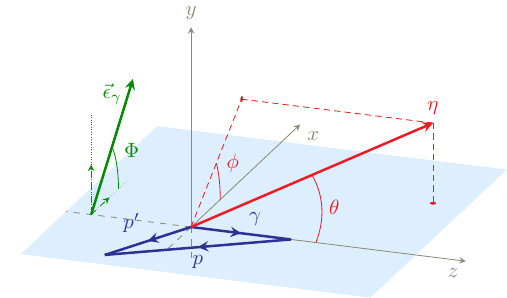}
\end{center}
\caption{\label{fig:frame}Definition of the angles in the Gottfried-Jackson frame.
In the two-meson rest frame, the $z$ axis is given by the photon beam ($\gamma$), and the  $xz$ reaction plane contains also the nucleon target ($p$) and recoiling nucleon ($p'$) momenta. $\theta$ and $\phi$ are the polar and azimuthal angles of the $\eta$. The polarization vector of the photon $(\vec{\epsilon}_\gamma)$ forms an angle $\Phi$ with the reaction plane.  Taken from Ref. \cite{Mathieu:2019fts}.
}
\end{figure}

We consider the photoproduction on a nucleon target of a meson resonance decaying into two spinless mesons, {\it e.g.} $\gamma p \to p\, \eta \, \pi^0$. We follow Ref.~\cite{Mathieu:2019fts}, writing
\begin{align} \nonumber
\label{eq:def_diffI}
\mathcal{I}(\Omega,\Phi) &= \frac{\diff\sigma}{\diff t \, \diff m_{\eta\pi^0} \, \diff\Omega \, \diff\Phi} \\
& = \kappa
\sum_{ \substack{\lambda_\gamma \lambda'_\gamma \\ \lambda_1\lambda_2}} A_{\lambda_\gamma; \lambda_1\lambda_2} (\Omega) \rho^\gamma_{\lambda_\gamma\lambda'_\gamma}(\Phi) A_{\lambda'_\gamma; \lambda_1\lambda_2}^* (\Omega),
\end{align}
where $\Omega = (\theta,\phi)$ are the decay angles of the resonance in the Gottfried-Jackson or helicity frame, and $\Phi$ is the polarization angle with respect to the production plane. The photon spin-density matrix is given by $\rho_\gamma(\Phi) = \frac{1}{2}\left(1 -P_{L} \cos 2\Phi\,\sigma_x - P_{L} \sin 2\Phi\,\sigma_y - P_{C}\sigma_z\right)$, and $P_{L}$ indicates the degree of linear polarization and $P_{C}$ circular. A full discussion and details for this are given in \cref{app:EllipticalPhotonSDM} and \cref{app:VirtualPhoton}.

The angular dependence is expanded in partial waves:
\begin{align}
A_{\lambda_\gamma; \lambda_1\lambda_2} (\Omega) = \sum_{\ell m} T^\ell_{\lambda_\gamma m; \lambda_1\lambda_2} Y^{m}_\ell(\Omega).
\end{align}
Parity conservation in the strong interaction reduces the number of production amplitudes $T^\ell_{\lambda_\gamma m; \lambda_1\lambda_2}$ by two. After the parity constraints have been implemented, the four nucleon helicity combinations reduce to two independent helicity difference amplitudes at the nucleon vertex. With an unpolarized target, these two combinations always appear as incoherent sums in observables. We consider only one nucleon transition in our simulation and suppress the nucleon helicities in the following derivations.\footnote{The interested reader will find the derivation and complete formulas in Ref.~\cite{Mathieu:2019fts}.} 

Finally, the photon helicity is transformed to the reflectivity basis :
\begin{align}  \label{def:Teps}
[\ell]^{(\epsilon)}_m &=  \frac{1}{2} \left[ T^\ell_{+1 m}  - \epsilon (-1)^m T^\ell_{-1 -m}\right].
\end{align}
The reflectivity component $\epsilon = \pm$ can be shown to correspond to the exchange of mesons of spin $J$ and parity $P = \epsilon (-1)^J$ in the high-energy limit~\cite{Mathieu:2019fts}. The reflectivity basis effectively decomposes the production amplitudes into good naturality Regge exchanges. 

For convenience, we define the amplitudes $U^{(\epsilon)}$ and $\tilde U^{(\epsilon)}$ in the
reflectivity basis:
\bsub\label{eq:def_U}\begin{align}
    U^{(\epsilon)}(\Omega) & = \sum_{\ell m} [\ell]^{(\epsilon)}_m Y^m_\ell(\Omega), \\
    \tilde U^{(\epsilon)}(\Omega) & = \sum_{\ell m} [\ell]^{(\epsilon)}_m \left[Y^m_\ell(\Omega) \right] ^*,
\end{align}\esub

\color{black}

We write the intensity of the final products from \cref{eq:def_diffI},
\begin{align}\label{eq:def_Ipre}
    \mathcal{I}(\Omega,\Phi) = \mathcal{I}^0(\Omega)
    &-P_{L} \mathcal{I}^1(\Omega)\cos(2\Phi)\notag \\
    &-P_{L} \mathcal{I}^2(\Omega) \sin(2\Phi)
    -P_{C} \mathcal{I}^3(\Omega),
\end{align}
where $\mathcal{I}^\alpha$ : $\mathcal{I}^0$ is the unpolarized; $\mathcal{I}^{1,2}$ are linearly polarized; and $\mathcal{I}^3$ the circularly polarized intensities. 
We only considered a polarized beam. In the case of a polarized target (and/or recoil), additional intensities can be extracted. The complete list is provided in Ref.~\cite{Roberts:2004mn}.
The intensities are bilinear in the partial waves and can be expressed in terms of the amplitudes in \cref{eq:def_U}:
\begin{align}\label{eq:def_I}
  \mathcal{I}^0(\Omega)& = \phantom{-2}\kappa \sum_\epsilon \left\{|U^{(\epsilon)}(\Omega)|^2 + |\tilde U^{(\epsilon)}(\Omega)|^2 \right\}\, ,\notag\\
  \mathcal{I}^1(\Omega)& = -2\kappa \sum_\epsilon \epsilon \re\left\{ U^{(\epsilon)}(\Omega)\, \left[\tilde U^{(\epsilon)}(\Omega)\right]^* \right\} \, ,\notag\\
  \mathcal{I}^2(\Omega)& = -2\kappa\sum_\epsilon \epsilon \im\left\{ U^{(\epsilon)}(\Omega)\, \left[\tilde U^{(\epsilon)}(\Omega)\right]^* \right\} \, ,\notag\\
  \mathcal{I}^3(\Omega)& = \phantom{-2}\kappa\sum_\epsilon \left\{|U^{(\epsilon)}(\Omega)|^2 - |\tilde U^{(\epsilon)}(\Omega)|^2 \right\}\, . 
\end{align}

In \cite{Mathieu:2019fts} the relationship between the meson spin density matrix elements (SDME) and the spherical harmonic components of the intensity in Eq.~\eqref{eq:def_Ipre} was derived, and the intensity expanded into a complete basis of the meson decay angles:
\begin{align} \label{eq:IntensityMoments}
    \mathcal{I}^\alpha(\Omega) = \kappa \sum_{L,M} H^\alpha(LM) Y^m_\ell(\Omega), 
\end{align}
The $H^\alpha(LM)$ are the moments of the spherical harmonic distributions $Y^m_\ell(\Omega)$, which contain all the decay information that can be extracted from the data. Moments are given in terms of the SDMEs, $\rho^{\alpha,\ell\ell'}_{mm'}$ of the reaction by
\begin{align} \label{eq:H_rho}
    H^\alpha(LM) = \sum_{\ell\ell',mm'}  \left( \frac{2\ell'+1}{2\ell+1} \right) ^{\frac{1}{2}}  C^{\ell 0}_{\ell' 0L0}  C^{\ell m}_{\ell' m'LM} \rho^{\alpha,\ell\ell'}_{mm'}.
\end{align}
Due to the property of the reflectivity basis, the two $\epsilon = \pm$ components are added incoherently to SDMEs:
\begin{align} \label{eq:reflSDME}
    \rho^{\alpha,\ell\ell'}_{mm'} = \hspace{0.2cm}^{+}\rho^{\alpha,\ell\ell'}_{mm'} \hspace{0.2cm}+\hspace{0.2cm} ^{-}\rho^{\alpha,\ell\ell'}_{mm'},
\end{align}
which are given by
\begin{align} \label{eq:rho_pws}
    ^{\epsilon}\rho^{0,\ell\ell'}_{mm'} =&
        [\ell]^{(\epsilon)}_{m}[\ell]^{(\epsilon)*}_{m'}
        +(-1)^{m-m'}[\ell]^{(\epsilon)}_{-m}[\ell]^{(\epsilon)*}_{-m'},\notag\\
    ^{\epsilon}\rho^{1,\ell\ell'}_{mm'} =& -\epsilon \left(
        (-1)^{m}[\ell]^{(\epsilon)}_{-m}[\ell]^{(\epsilon)*}_{m'}
        +(-1)^{m'}[\ell]^{(\epsilon)}_{m}[\ell]^{(\epsilon)*}_{-m'} \right), \notag\\
    ^{\epsilon}\rho^{2,\ell\ell'}_{mm'} =&-i\epsilon \left(
        (-1)^{m}[\ell]^{(\epsilon)}_{-m}[\ell]^{(\epsilon)*}_{m'}
        -(-1)^{m'}[\ell]^{(\epsilon)}_{m}[\ell]^{(\epsilon)*}_{-m'} \right),\notag\\
    ^{\epsilon}\rho^{3,\ell\ell'}_{mm'} =&
        [\ell]^{(\epsilon)}_{m}[\ell]^{(\epsilon)*}_{m'}
        -(-1)^{m-m'}[\ell]^{(\epsilon)}_{-m}[\ell]^{(\epsilon)*}_{-m'}.\notag\\
\end{align}

Where the complex partial waves are written as $[\ell]^{(\epsilon)}_m = |[\ell]^{(\epsilon)}_m | \exp(i\phi_{\ell,m}^{(\epsilon)})$. 

Previous work on single pseudoscalar meson photoproduction~\cite{Wunderlich:2013iga,Tiator:2017cde} has shown that the expanding the production amplitudes in a truncated set of partial waves provides a means of effectively performing a complete experiment : all other observables can be calculated from given results, with significantly fewer experimental observables. We observe a similar reduction here and associate it with the number of spherical harmonic moments that are accessible with polarized beams for a given $\ell_{max}$.
~\cref{App:counting} shows how the number of partial waves and measurable moments increases with $\ell_{max}$ for the case of linearly and elliptically polarized measurements. It is a necessary condition that the number of experimentally measurable moments exceeds the number of partial waves to extract the partial waves, which is clearly met in the general case shown.

\begin{table}[tbp]
    \centering
    \caption{Number of waves; real numbers; SDMEs; Moments for $\alpha=0,1,2$ and $\alpha=0,1,2,3$. Formula are given in ~\cref{App:counting}.}
     \label{tab:counting}
    \begin{tabular}{|c|c|c|c|c|c|}
    \hline
    $\ell_{max}$ & $N_{waves}$ & $N_{real}$ & $N_{sdme}$ & $N_{m012}$ & $N_{m0123}$ \\
    \hline
    0 & 2 & 2 & 3 & 2 & 2 \\
    \hline
    1 & 8 & 14 & 36 & 15 & 18 \\
    \hline
    2 & 18 & 34 & 171 & 40 & 50 \\
    \hline
    3 & 32 & 62 & 528 & 77 & 98 \\
    \hline
    4 & 50 & 98 & 1275 & 126 & 162 \\
    \hline
    \end{tabular}
\end{table}

In Guo et al. \cite{kjcp-h8b9}, Appendix A, it was shown after substituting equations \cref{eq:rho_pws} into \cref{eq:H_rho} the moments $H^\alpha(LM)$ for $\alpha=0,1,2$ depend on the sum of terms given by the real parts of $\rho^{\alpha,\ell\ell'}_{mm'}$, related to $\cos{\phi_{\ell,m}^{(\epsilon)}}$. This leads to the fact that if we only measure the $\alpha=0,1,2$ intensities, we can only find the magnitude of the partial-wave phases up to an ambiguous sign. That is, we will always have two complex conjugate solutions, which is consistent with the findings of \cite{PhysRevD.108.076001}. It is clear, however, if we can also measure the imaginary parts we will resolve these ambiguities. In \cref{app:h3_derivation} we show that the $H^3$ moments coming from the circular polarization, so indeed give access to the imaginary part of the partial wave, thereby resolving this complex conjugate ambiguity. Specifically $H^3$ moments can be written in terms of the sum of terms~:
\begin{equation*}
    \im\left([\ell]_{m}^{(\epsilon)}[\ell^{\prime}]_{m^{\prime}}^{(\epsilon)*}\right) =  |[\ell]_{m}^{(\epsilon)}|| [\ell^{\prime}]_{m^{\prime}}^{(\epsilon)}|\sin(\phi_{\ell,m}^{(\epsilon)}-\phi_{\ell',m'}^{(\epsilon)}) 
\end{equation*}
for $m\ne0$. 
In \cite{Mathieu:2019fts} it was shown due to parity constraints that the moments $H^{0,1}$ are real, while $H^{2,3}$ are imaginery. We can see how this relates to the current discussion by inspecting ~\cref{eq:rho_pws}: the $^{\epsilon}\rho^{\alpha,\ell\ell'}_{mm'}$ SDMEs for $\alpha=0,1$ are related to the real parts of the bilinear combinations of partial waves $[\ell]^{(\epsilon)}_{m}[\ell]^{(\epsilon)*}_{m'}$, as shown in \cite{kjcp-h8b9}, and so directly yield real moments; for $\alpha=2$ the SDMEs are also given by the real parts of  $[\ell]^{(\epsilon)}_{m}[\ell]^{(\epsilon)*}_{m'}$, but the SDME has an additional factor of $i$, making the SDMEs and therefore moments imaginary; the $\alpha=3$ case are directly related to the imaginary part of $[\ell]^{(\epsilon)}_{m}[\ell]^{(\epsilon)*}_{m'}$ and with no further factor of $i$ we get imaginary SDMEs and moments. In~\cref{App:Spin1H}, we show explicitly for the $\ell=1$ case, that the $H^{0,1,2}$ moments indeed involve only the cosine of phase differences, whereas the $H^3$ moments involve only sine of phase differences.




\subsection{Extracting Partial Waves from Moments of Spherical Harmonics}
Moments can be unambiguously extracted from the data, but the physics of interest is encoded in the partial waves.
The main tool of this analysis is a numerical procedure to solve the set of simultaneous equations that relate spherical harmonic moments to partial waves. These equations are generated in the general case by substituting \cref{eq:rho_pws} into \cref{eq:H_rho}. The moments can be extracted from the experimental data from the relative contributions of each spherical harmonic to the decay angle distribution. Typically, this is done by maximum likelihood fitting. In the current work, we start from amplitudes, use these to calculate the moments, and then solve the simultaneous equations to check consistency and ambiguities. 
As the equations are non-linear, they contain $\sin$ and $\cos$ terms and must be solved numerically. As we will see, it is not sufficient to just have more equations than unknowns, due to the dependencies and non-linearities within the system. Here we constructed a $\chi^2$ metric from the sum of the square of the differences between the
moment values, $\hat{H}^{\alpha}(L,M)$ and the evaluated moment equations given the partial wave components, $[\ell]_{m}^{(\epsilon)}$ : $H^{\alpha}(L,M)([\ell]_{m}^{(\epsilon)})$.

As we are inverting precise numerical values we do not account for uncertainties which would arise when dealing with real experimental data. Our $\chi^2$ simply reads:
\begin{equation}
    \chi^2 = \sum_{\alpha,L,M} \left( \hat{H}^{\alpha}(L,M) - H^{\alpha}(L,M)([\ell]_{m}^{(\epsilon)})\right)^2
\end{equation}

This was minimized using the Minuit2 package from CERN ROOT \cite{James:1975dr,Brun:1997pa}. The minimization was performed many times for each set of moments and the minimized values were recorded. For each minimization, a random set of partial waves was generated for starting values. Initially, random magnitudes and phases were drawn from flat distributions with the magnitudes then normalized so that the sum of their squares was equal to one. Often, minimization fails before reaching the global minimum or stops at a local minimum. Here we just kept the results of minimizations consistent with the lowest global $\chi^2$. When we generate moments directly from partial waves, the global $\chi^2$ equals zero within the precision of the computation.

\section{Revisiting case : S and D waves}
\label{sec:SDwaves}
In the work of Smith et al.~\cite{PhysRevD.108.076001} it was shown that in the general case there should be no ambiguity, other than a complex conjugate solution, when we have linearly polarized photoproduction. This was demonstrated with a specific example, which used a simplified set of S and D waves of a single reflectivity: $^+S$; $^+D_{-1}$; $^+D_{0}$ and $^+D_{1}$.
Thus, there are seven real numbers to extract from a maximum of 22 moments in the case of general photon polarization including all moments $H^\alpha(LM)$ with $\alpha = 0,1,2,3$. Including the second reflectivity would double the number of real numbers.

Here we re-analyze this case, but using the numerical inversion of the moment equations. Further, we explore the situation for adding different polarizations, including circular.

\subsection{Unpolarized}
Here we consider the case where we have no polarization information in the data and only have access to the $H^0$ moments, There are seven non-zero moments : $H^0(0,0), H^0(2,0), H^0(2,1), H^0(2,2), H^0(4,0), H^0(4,1),$ and $H^0(4,2)$.

\begin{figure*}[tbp]
    \centering
    \includegraphics[width=\textwidth]{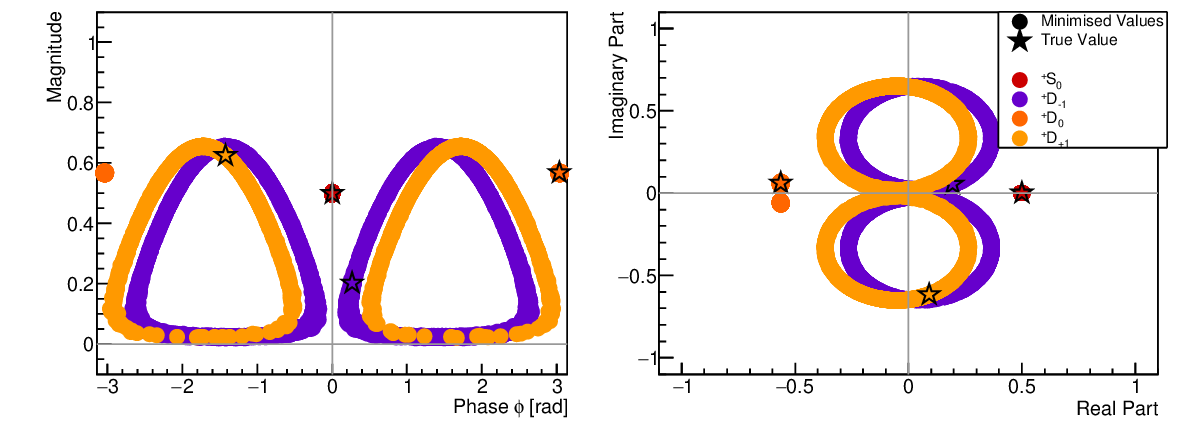}
    \caption{
        Results of the numerical inversion for the positive-reflectivity S- and D-wave amplitudes, using only unpolarized moments $H^{0}$. 
        The colored circles represents the results with minimal $\chi^{2}$ for 100k numerical inversions.
        The outlined star indicates the single true value used to generate the data for each partial wave.
        The left panel displays the magnitude versus phase, while the right panel shows the corresponding Argand diagrams (imaginary vs. real parts).
    }
    \label{fig:sd_wave_H0}
\end{figure*}

The results of the numerical inversion are illustrated in \cref{fig:sd_wave_H0}. The findings reveal a complex ambiguity structure. The solutions for the $S_0^+$ amplitude are found to be unique, while the $D_0^+$ amplitude exhibits a two-fold complex conjugate ambiguity. In contrast, the solutions for the $D_{+1}^+$ and $D_{-1}^+$ amplitudes are not discrete points but trace continuous ellipses on the Argand diagram.

This behavior is expected from the analytical properties of the moment equations and the conclusions of ~\cite{PhysRevD.108.076001}. As discussed in that work, when only unpolarized moments are used, the system is under-determined. For the seven unknown parameters (four magnitudes and three relative phases), there are only six independent moment equations, as $H^0(2,2) \propto H^0(4,2)$, which necessarily leads to a ``1-parameter continuous transformation" that leaves all moments invariant. The ellipses observed for the $D_{+1}^+$ and $D_{-1}^+$ amplitudes are the direct geometric manifestation of this continuous ambiguity; they represent the projection of the one-dimensional solution curve onto the complex planes of these amplitudes.

\subsection{Linearly Polarized - $\cos{2\Phi}$ only}
\begin{figure*}[tbp]
    \centering
    \includegraphics[width=\textwidth]{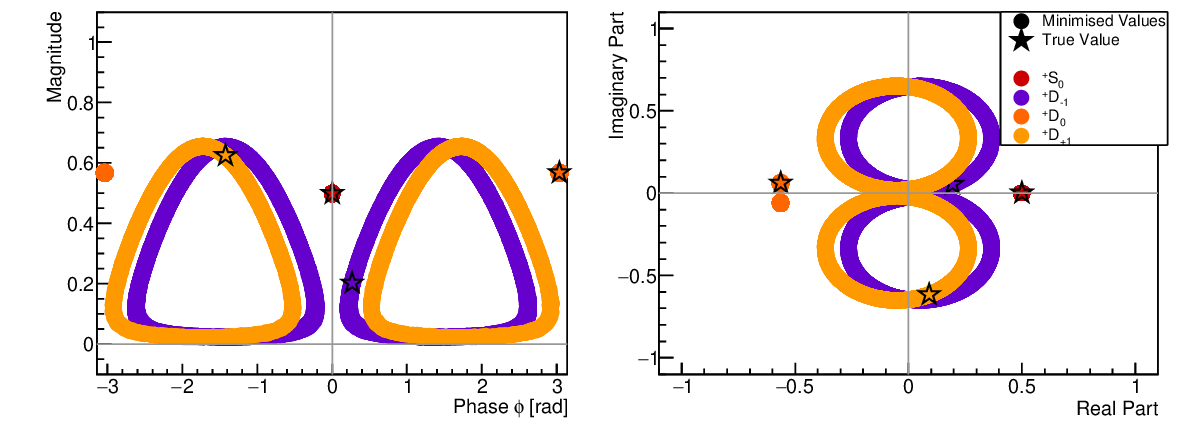}
    \caption{
        Results of the numerical inversion for the positive-reflectivity S- and D-wave amplitudes, including moments from both $H^{0}$ and ($H^{1}$). 
        The inclusion of the $H^{1}$ moments provides no additional constraints on the fit, leading to results that are identical to the $H^{0}$-only case shown in Figure~\ref{fig:sd_wave_H0}. 
        As before, the colored circles represent the best $\chi^{2}$ solutions from 100k numerical inversions, and the outlined star indicates the true value for each partial wave.
    }
    \label{fig:sd_wave_H0H1}
\end{figure*}

The results of including the equations for the moments $H^0$ and $H^1$ in the numerical inversion are shown in ~\cref{fig:sd_wave_H0H1}. Here we clearly see that there is no change from the case of $H^0$ moments only.
This can be understood directly from the structure of the moment equations. The continuous ambiguity, which manifests as elliptical solution paths for the $D_{+1}$ and $D_{-1}$ amplitudes, is fundamentally tied to the inability to uniquely determine their individual magnitudes, $|D_{+1}^{+}|$ and $|D_{-1}^{+}|$. Although the $H^1(2,2)$ moment provides a clean constraint on the \textit{sum} of their squared magnitudes, $|D_{+1}^{+}|^2 + |D_{-1}^{+}|^2$, it does not provide information on their \textit{difference}. This crucial missing piece of information is provided by the $H^2(2,2)$ moment, which is directly proportional to $|D_{-1}^{+}|^2 - |D_{+1}^{+}|^2$. Without this second independent constraint, the system remains under-determined with respect to these two magnitudes, and the continuous ambiguity must persist.

\subsection{Fully Linearly Polarized}
\begin{figure*}[tbp]
    \centering
    \includegraphics[width=\textwidth]{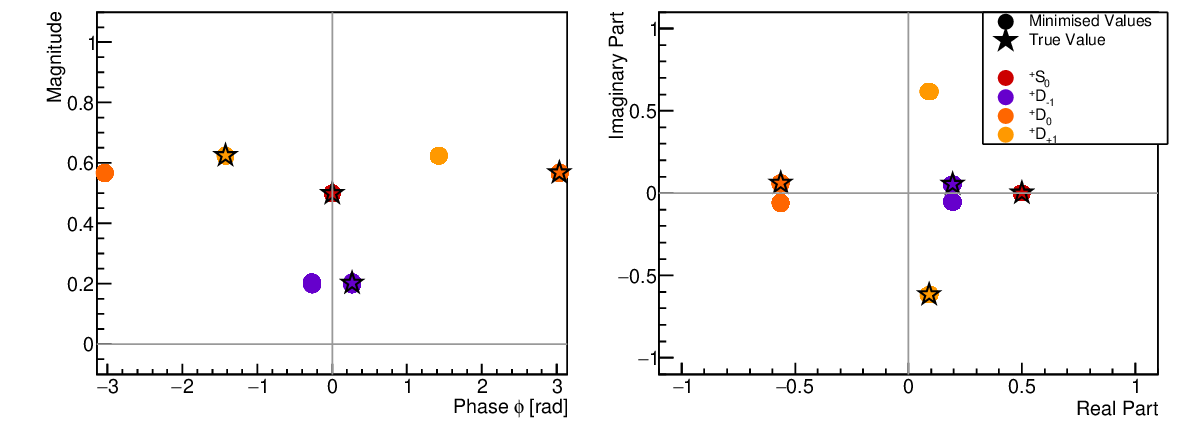}
    \caption{
        Results of the numerical inversion after including moments $H^{2}$, in addition to $H^{0}$ and $H^{1}$.
        The constraints from the $H^{2}$ moments successfully resolve the continuous ambiguities observed in previous fits (cf. Figure~\ref{fig:sd_wave_H0H1}). 
        The fit now converges to only two discrete solutions: the true value and its trivial complex conjugate ambiguity.
        The plot shows the minimised fit result (solid circles) converging exactly onto the true generated values (outlined stars).
    }
    \label{fig:sd_wave_H0H1H2}
\end{figure*}

Here we also include the $H^2$ moment equations to give us a fully linearly polarized experiment. The results are shown in ~\cref{fig:sd_wave_H0H1H2} As expected from the previous section, we now resolve the continuous ambiguity via the $H^2(2,2)$ moment. We are left with the complex conjugate ambiguity as already determined in \cite{PhysRevD.108.076001}.
\subsection{Circularly Polarized}
Another interesting case arises when one considers an experiment with only unpolarized ($H^0$) and circularly polarized ($H^3$) beams. This combination of moments provides constraints on both the real (cosine) and imaginary (sine) parts of the interference terms, which is sufficient to resolve the simple complex conjugate ambiguity. In addition we have 9 different moment equations (11 non-zero with $H^\alpha(2,2)\propto H^\alpha(4,2)$), so in principle this may be enough to resolve all ambiguities. However, a different two-fold discrete ambiguity emerges. 
The numerical inversion of the system for the specific amplitudes considered here reveals a two-fold discrete ambiguity. This can be understood as a `near-symmetry' in the system. A detailed analysis shows that the moment equations are almost invariant under a transformation of the type $D_{+1} \leftrightarrow D_{-1}^*$, $D_0 \leftrightarrow D_0^*$, etc. The symmetry is only broken by the moments with $M=1$, namely $H^0(2,1)$ and $H^0(4,1)$. For the Smith et al. values, these moments are small but nonzero ($H^0(2,1) = -0.041$ and $H^0(4,1) = 0.009$). This slight breaking of the symmetry `warps' the solution space, resulting in two distinct but numerically close solutions rather than a perfect degeneracy. This demonstrates that even in an overconstrained system, discrete ambiguities can arise for specific physical scenarios where the amplitudes lie close to a point of higher symmetry.

\begin{figure*}[tbp]
    \centering
    \includegraphics[width=\textwidth]{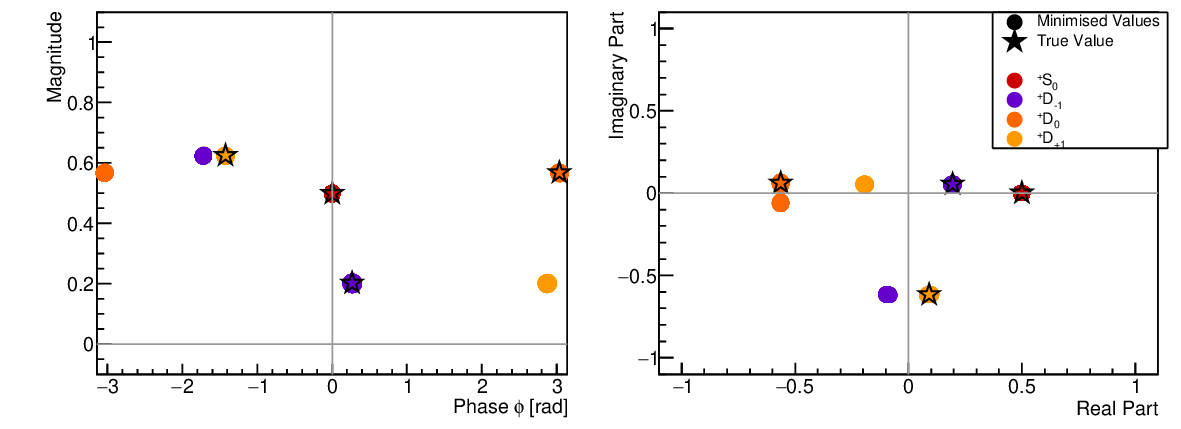}
    \caption{
        Results of the numerical inversion using moments ($H^{0}$) and ($H^{3}$), under the constraint that only a single reflectivity state is present in the model. 
        This specific set of constraints resolves the continuous ambiguity but results in a single, discrete ambiguity. 
        The plot compares the true generated values (outlined stars) with the two ambiguous fit solutions. The first solution is shown as solid circles, while the second, discrete ambiguous solution is indicated by the colored dashed lines.
        The left panel displays the magnitude versus phase, while the right panel shows the corresponding Argand diagrams for the positive-reflectivity amplitudes.
    }
    \label{fig:sd_wave_H0H3}
\end{figure*}
\subsection{Elliptically Polarized}

\begin{figure*}[tbp]
    \centering
    \includegraphics[width=\textwidth]{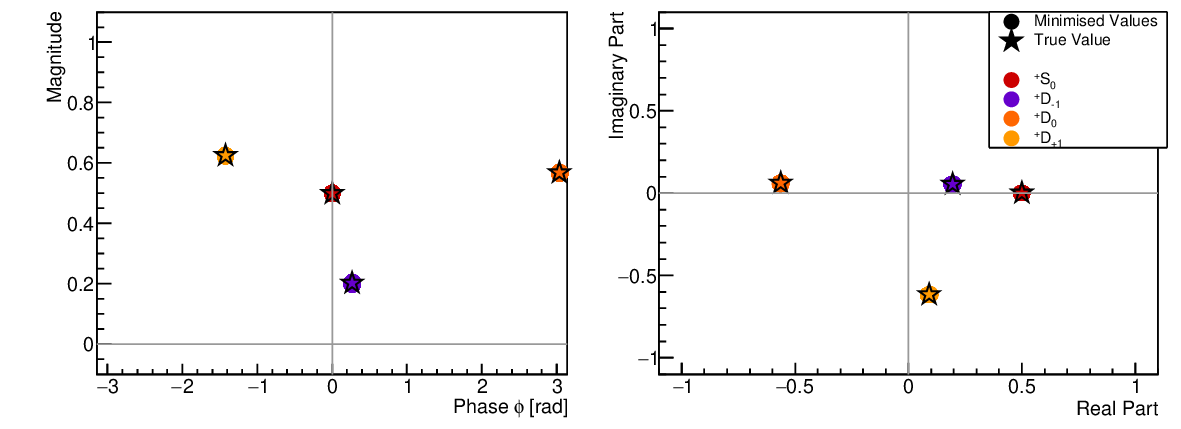}
    \caption{
        Results of the numerical inversion after the inclusion of all moments up to $H^{3}$. 
        The addition of the $H^{3}$ moments provides the final constraints necessary to resolve all mathematical ambiguities in the fit.
        This removes the complex conjugate ambiguity that remained in the $H^{0,1,2}$-only fit (cf. Figure~\ref{fig:sd_wave_H0H1H2}), leading to a single, unique solution.
        The plot demonstrates this outcome, showing the minimised fit result (solid circles) converging perfectly on the true generated values (outlined stars).
    }
    \label{fig:sd_wave_H0H1H2H3}
\end{figure*}

Now, combining all possible photon polarizations to have an elliptically polarized experiment, we see in \cref{fig:sd_wave_H0H1H2H3} that we fully resolve all possible ambiguities in this example. This is because we now have access to the imaginery part of the SDMEs, as derived in ~\cref{app:h3_derivation}.

\begin{figure*}[tbp]
    \centering
    \includegraphics[width=\textwidth]{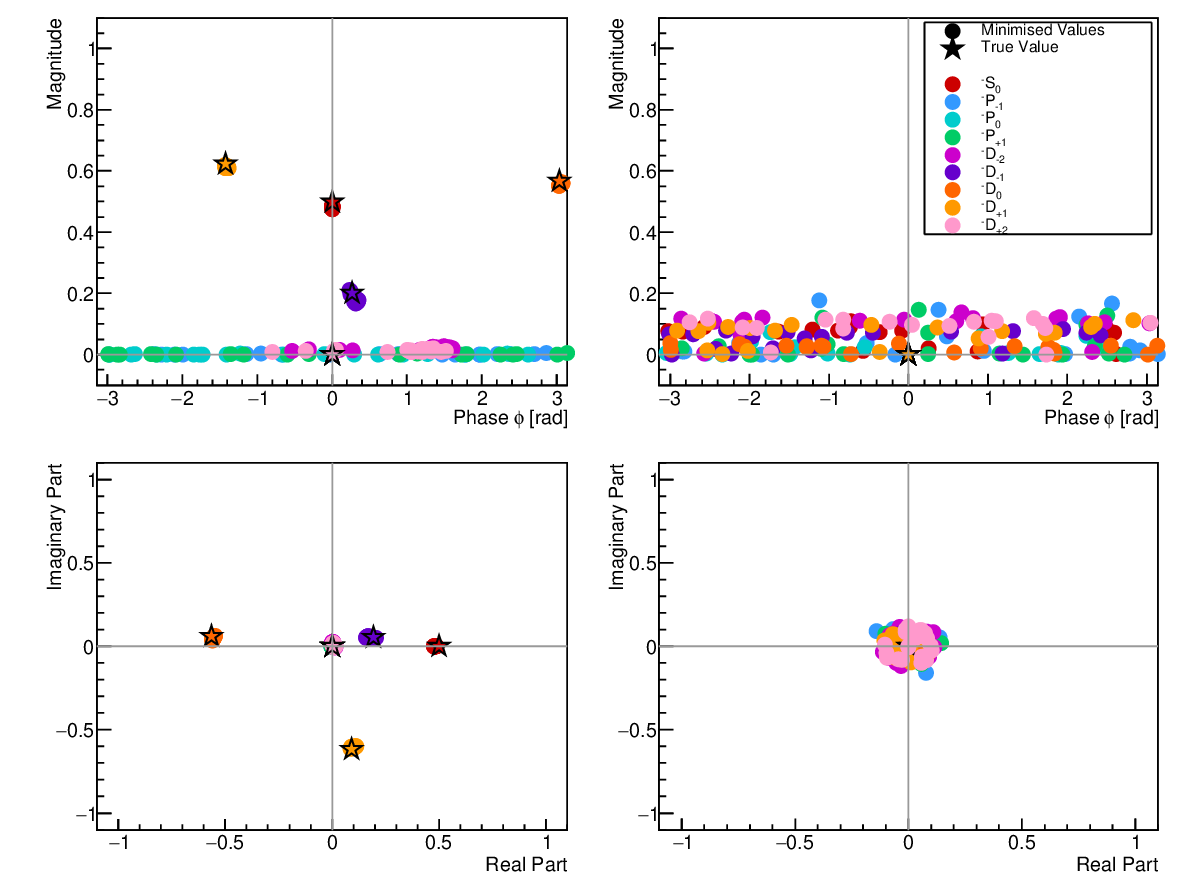}
    \caption{
        Results of the numerical inversion for the full basis of S, P, and D partial wave amplitudes, constrained by all moments up to $H^{3}$.
        The inclusion of moments derived from double-polarization observables ($H^{3}$) provides the final constraints needed to resolve all mathematical ambiguities, leading to a single, unique fit solution.
        The plot demonstrates this definitive result, showing the minimised fit solution (solid circles) converging perfectly onto the true generated values (outlined stars) for all partial waves.
        The top row displays the magnitude versus phase, while the bottom row shows the corresponding Argand diagrams. The left and right columns show the positive ($^{+}S_0$, etc.) and negative ($^{-}P_{-1}$, etc.) reflectivity states, respectively.
    }
    \label{fig:sd_waves_h0123_2refl}
\end{figure*}

Including P-waves and the negative reflectivity in the numerical inversion procedure does not have an adverse effect on the results. This is as expected from the work of Smith et al.~\cite{PhysRevD.108.076001}, and is confirmed with the numerical technique. If we also add in the $m = 2$ $D$-waves and double the waves by including the negative reflectivity, then again we only find single solutions for the case of elliptically polarized and a single complex conjugate ambiguity in the case of linear polarization only. This is shown for the elliptical case in ~\cref{fig:sd_waves_h0123_2refl}. While the zero positive reflectivity waves are minimized to zero, the negative reflectivity waves show some noise,  fluctuating to magnitude up to 0.1, which corresponds to $1\%$ in intensity, probably due to the finite minimization precision.

In ~\cref{App:logVal} the amplitude versus $\chi^2$ parameter space is investigated. We see that inclusion of elliptical polarization gives greater separation between the fake minima and the true one. In addition, the additional constraints will provide smaller statistical uncertainty, particularly if the degree of circular polarization is greater than linear. These aspects point towards elliptical polarization having a practical benefit in partial wave analysis.
 
\section{Revisiting case : $P$-Waves with no $m=0$ Component} \label{Sec:pwavem=0}
In this Section we look again at Section IV:B from Guo et al. \cite{kjcp-h8b9}. There they looked at a simple wave set which includes just the $P$-waves with $m$ projections $\pm1$ and both reflectivities. They found that this special case resulted in a continuous ambiguity and demonstrated that this was a result of the equations relating moments to partial-waves simplifying to 5 constraints for the 6 real amplitude parameters. This directly aligns with the criterion for continuous ambiguities where the number of parameters ($N_{par}=6$) exceeds the rank of the linear equations ($rank(A)=5$).

As in Section \cref{sec:SDwaves}, we solve the simultaneous equations numerically. We use the same amplitude values as \cite{kjcp-h8b9} as shown in \cref{tab:input_pwave},
\begin{table}[h!]
\centering
\caption{Input P-wave amplitude values for the numerical example. The magnitude of $P_{+1}^{+}$ is derived from the constraint $\sum|P|^2=1$.}
\begin{tabular}{|c|c|c|}
\hline
\textbf{Amplitude} & \textbf{Magnitude} & \textbf{Phase (radians)} \\
\hline
$P_{+1}^{+}$ & 0.860 & 0.0 (fixed) \\
$P_{-1}^{+}$ & 0.130 & $-1.114$ \\
$P_{+1}^{-}$ & 0.142 & 0.0 (fixed) \\
$P_{-1}^{-}$ & 0.473 & $-1.635$ \\
\hline
\end{tabular}
\label{tab:input_pwave}
\end{table}

\begin{figure*}[tbp]
\centering
\includegraphics[width=\textwidth]{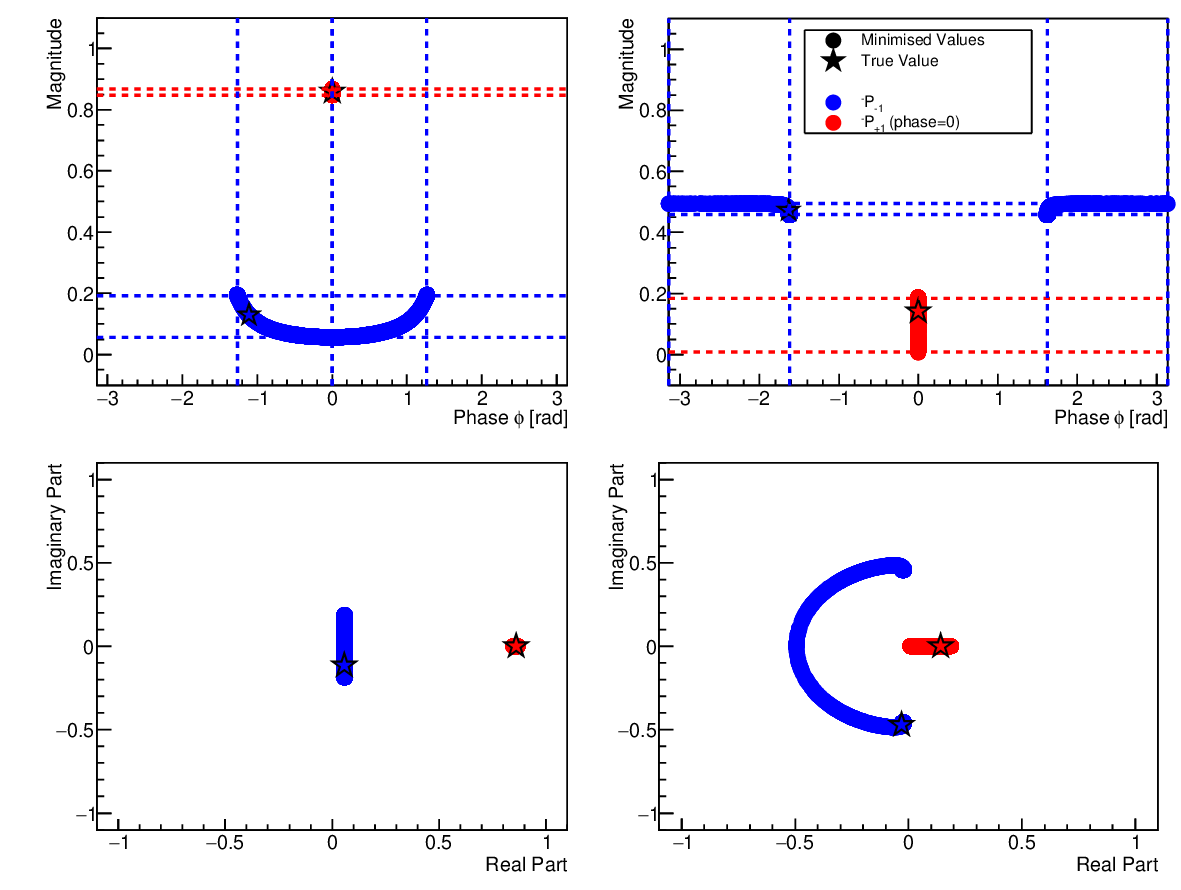}
\caption{\label{fig:PWave_H012}
Comparison of minimised partial wave amplitudes against their true values for a fit including H0, H1, and H2 components. 
The solid colored circles represent the results from the fit solutions with the lowest log-likelihood values, illustrating a continuous ambiguity. 
The outlined stars indicate the single set of true, generated values. 
The dashed lines show the calculated allowed ranges which bound this ambiguity for each corresponding partial wave as explained in \cref{app:p_wave_constraints}. 
The top row displays magnitude versus phase, while the bottom row shows the corresponding Argand diagrams (imaginary vs. real parts). 
Left and right panels correspond to positive ($^{+}P_{\pm 1}$, etc.) and negative ($^{-}P_{\pm 1}$, etc.) reflectivity states, respectively.
}
\label{fig:pwa_h0h1h2_ambiguity}
\end{figure*}

The results of the numerical inversion are shown in \cref{fig:PWave_H012}.

The results clearly illustrate the continuous ambiguity which was found. In addition here we derive values which define the ranges the parameters can have. This is shown in detail in \cref{app:p_wave_constraints}. The procedure to determine the allowed parameter ranges involves four main steps. First, the five independent, non-zero moments are transformed into a set physical observables: the sum of intensities ($I^\epsilon$), the difference of intensities ($M^\epsilon$), and the real part of the interference term ($R^\epsilon$).  
The physical observables are given by,
\begin{align}
\begin{split}
        I^{\epsilon} &= |P_{+1}^{\epsilon}|^2 + |P_{-1}^{\epsilon}|^2 \\
        M^{\epsilon} &= |P_{-1}^{\epsilon}|^2 - |P_{+1}^{\epsilon}|^2 \\
        R^{\epsilon} &= |P_{+1}^{\epsilon}||P_{-1}^{\epsilon}|\cos(\phi_{P_{-1}}^{\epsilon})
\end{split}
\end{align}
and these relate to measured moments by,
\begin{align}
\begin{split}
        I^{+} + I^{-} &= \frac{1}{2}H_{0}(0,0) \\
        I^{+} - I^{-} &= \frac{5}{\sqrt{6}}H_{1}(2,2) \\
        R^{+} &= -\frac{1}{8} \left( \frac{10}{\sqrt{6}}H_{0}(2,2) + H_{1}(0,0) \right) \\
        R^{-} &= \frac{1}{8}H_{1}(0,0) - \frac{5}{4\sqrt{6}}H_{0}(2,2) \\
        M^{+} - M^{-} &= \frac{5}{i\sqrt{6}}H_{2}(2,2)
\end{split}
\end{align}

Second, using the fundamental constraint $|\cos(\phi)| \le 1$, a preliminary, symmetric range for the magnitudes is derived for each reflectivity sector independently. Third, the ambiguity range from one reflectivity sector is used to constrain the other via the linking observable ($M^+ - M^-$), yielding a new, asymmetric set of ranges. Finally, the true allowed range for each amplitude is found by taking the intersection of the ranges derived in the second and third steps, after having enforced the physical requirement that all intensities must be positive (i.e. $|M^\epsilon| \le I^\epsilon$). For a complete derivation see ~\cref{app:p_wave_constraints}.

We see that the calculated ranges agree precisely with those found from the numerical inversion. This highlights that analysis of the Moments-PartialWaves equations can give insight into the results we can extract from data.

\subsection{Addition of Circular Polarization}
Now we fully understand the presence of the continuous ambiguity we can consider the additional constraints coming from the circularly polarized moments. As the $H^0, H^1, H^2$ moments only constrain the real parts of the bilinear amplitude products, they are susceptible to this ambiguity. In contrast, the circularly polarized moment $H^3$ provides access to the imaginary parts of these products. In this simplified case, we have one additional non-zero moment, $H^{3}(2,2)$, which provides the necessary additional constraint:
\begin{align}
    H^{3}(2,2)/i =\frac{2\sqrt{6}}{5} ( &|P_{+1}^{+}||P_{-1}^{+}|\sin(\phi_{P_{+1}}^{+} - \phi_{P_{-1}}^{+}) \notag\\
    +&|P_{+1}^{-}||P_{-1}^{-}|\sin(\phi_{P_{+1}}^{-} - \phi_{P_{-1}}^{-}) )
\end{align}

\begin{figure*}[tbp]
\centering
\includegraphics[width=\textwidth]{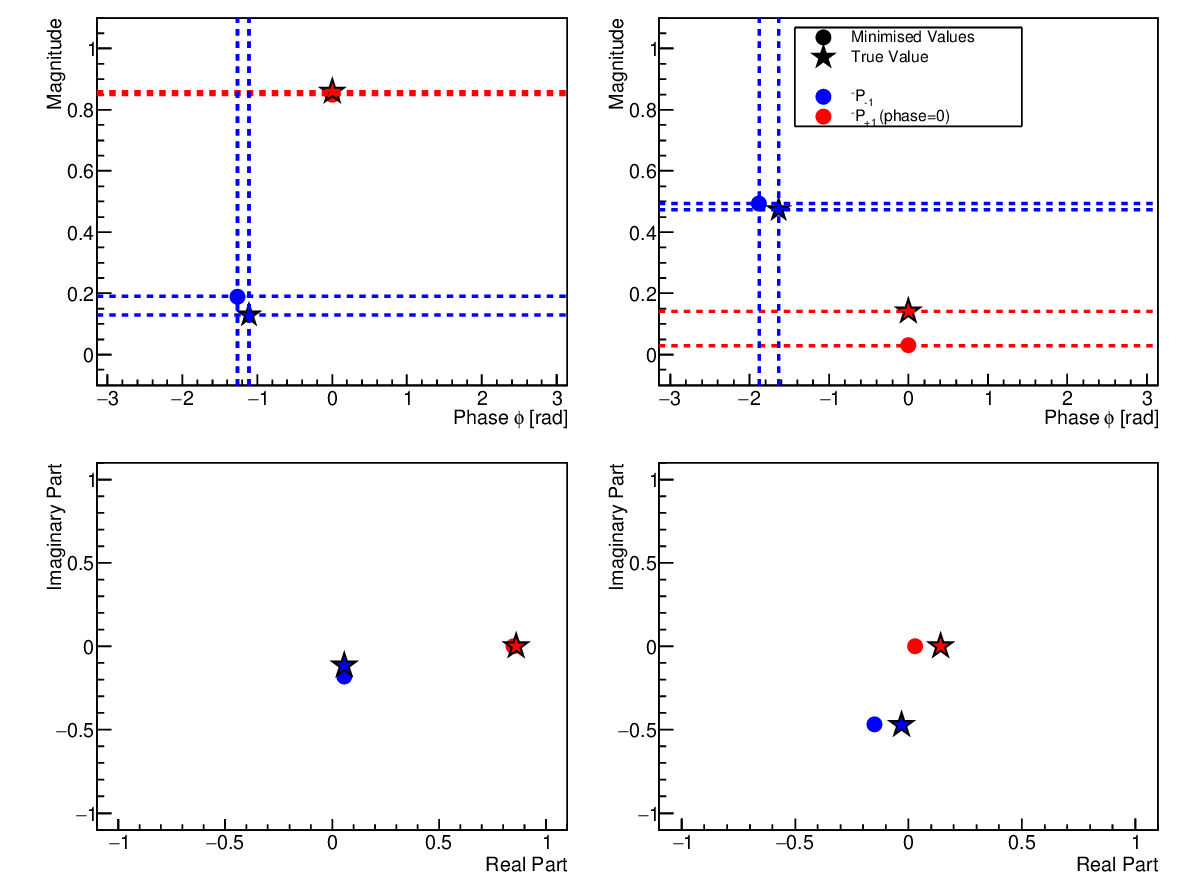}
\caption{
Comparison of the two discrete ambiguous fit solutions against the single true value for a fit including $H^{0,1,2,3}$ moments. 
The solid colored circles represent the results from one of the numerical inversion, while the dashed lines indicate the corresponding values for the calculated discrete ambiguous solutions. 
The outlined stars show the single set of true, generated values. 
The top row displays magnitude versus phase, and the bottom row shows the corresponding Argand diagrams. 
Left and right panels correspond to positive ($^{+}P_{\pm 1}$, etc.) and negative ($^{-}P_{\pm 1}$, etc.) reflectivity states, respectively.
}
\label{fig:pwa_h0h1h2h3_discrete}
\end{figure*}

In \cref{app:resolving_ambiguity} we show how this condition allows us to avoid the continuous ambiguity that persists with only linear polarization in the simplified $\{P_{+1}^{\pm}, P_{-1}^{\pm}\}$ model. The $H^3(2,2)$ moment, determines the sum of the imaginary parts of the interference terms ($S^+ + S^-$). This new information, when combined with the fundamental trigonometric identity $\sin^2\phi + \cos^2\phi = 1$, creates a fully determined, albeit non-linear, system of equations. This system no longer permits a continuous family of solutions but instead yields a finite number of discrete solutions. The continuous ambiguity is resolved into a two-fold discrete ambiguity. This arises from solving a system of quadratic equations which, geometrically, corresponds to the intersection of two ellipses and thus, in this case, yields two real solutions.
The numerical example in Appendix~\ref{app:resolving_ambiguity} corresponds to the current wave set and the lines on \cref{fig:pwa_h0h1h2h3_discrete} now correspond to the 2 actual solutions, rather than limits of ambiguous solutions; while the data points show the true and inverted solutions as before. This ambiguity is distinct from the trivial complex conjugate ambiguity and is a direct consequence of the non-linear nature of the system.

\section{Polarized $\rho$ photoproduction} \label{Sec:pol_rho}
\subsection{Spin Density Matrix Elements}

In this section we examine $\rho$ photoproduction as the SDMEs have recently been measured by the GlueX collaboration \cite{PhysRevC.108.055204} providing a realistic set of amplitudes. In addition this reaction is dominant and therefore produces a relatively large volume of events compared to other reactions. Hence polarized $\rho$ photoproduction makes a promising candidate for the polarimetry applications suggested in this article.

In terms of the SDMEs $\rho^\alpha_{\lambda_{V}\lambda_{V'}}$, with $\lambda_{V}$ the vector meson helicity, as defined in \cite{SCHILLING1970397} the polarized intensities of \cref{eq:def_Ipre} are written as,

\begin{align}\label{eq:IrhoSDMEs}
    \mathcal{I}_{0}(\Omega)=&\frac{3}{4\pi}\{
    \frac{1}{2}(1-\rho^{0}_{00}) 
    +\frac{1}{2}(3\rho^{0}_{00} - 1)\cos^{2}{\theta}\notag\\
    &-\sqrt{2}\Re{[\rho^{0}_{10}]}\sin{2\theta}\cos{\phi}
    -\rho^{0}_{1-1}\sin^{2}{\theta}\cos{2\phi}
    \},\notag\\
    \mathcal{I}_{1}(\Omega)=&\frac{3}{4\pi}\{
    \rho^{1}_{11}\sin^{2}{\theta} + 
    \rho^{1}_{00}\cos^{2}{\theta}\notag\\ 
    &-\sqrt{2}\rho^{1}_{10}\sin{2\theta}\cos{\phi}
    -\rho^{1}_{1-1}\sin^{2}{\theta}\cos{2\phi}
    \},\notag\\
    \mathcal{I}_{2}(\Omega)=&\frac{3}{4\pi}\{
    \sqrt{2}\Im{\rho^{2}_{10}}\sin{2\theta}\sin{\phi}
    +\Im{\rho^{2}_{1-1}}\sin^{2}{\theta}\sin{2\phi}
    \},\notag\\
    \mathcal{I}_{3}(\Omega)=&\frac{3}{4\pi}\{
    \sqrt{2}\Im{\rho^{3}_{10}}\sin{2\theta}\sin{\phi}
    +\Im{\rho^{3}_{1-1}}\sin^{2}{\theta}\sin{2\phi}
    \}.
\end{align}

GlueX published results at with linearly polarized photons at $E_{\gamma}=8.5$~GeV and covering a four-momentum transfer range of $-1<t/\text{GeV}^2<0$ with orders of magnitude more statistics than previous measurements. The results show that s-channel helicity conservation is dominant at small squared four-momentum transfer and extracted the t-dependence of natural
and unnatural-parity exchange contributions to the production process.

\begin{figure*}[tbp]
\centering
\includegraphics[width=\textwidth]{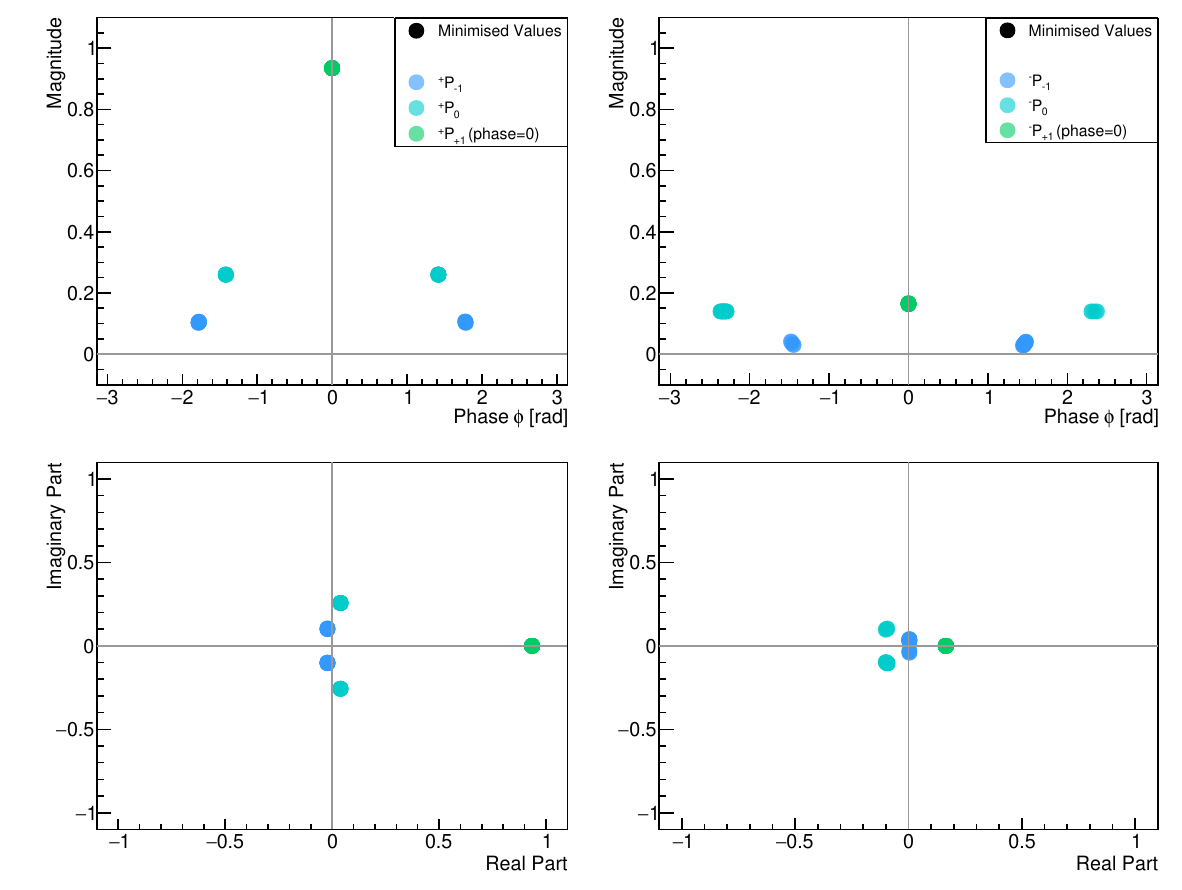}
\caption{
Resulting P-wave amplitudes from GlueX $\rho$ SDME data produced through inverting the equivalent $H^{0,1,2}$ moments. 
The solid colored circles represent the results from one of the numerical inversions. The top row displays magnitude versus phase, and the bottom row shows the corresponding Argand diagrams. 
Left and right panels correspond to positive ($^{+}P_{\pm 1}$, etc.) and negative ($^{-}P_{\pm 1}$, etc.) reflectivity states, respectively.
}
\label{fig:rho_sdme_t07}
\end{figure*}

The GlueX analysis assumed the P-wave contribution is dominant and therefore neglects smaller partial waves. Following the same assumption we deduced the P-wave contributions in terms of spin projections m, and reflectivty $\epsilon$. This can be done by first relating the $\rho$ SDMEs of \cref{eq:IrhoSDMEs} to the Spherical Harmonic moments defined by \cref{eq:IntensityMoments}. The resulting expressions are given in \cref{App:RhoSDMEMoments}. 
We choose GlueX data for at $t=-0.7278\text{ GeV}^2$ to produce the set of moments $H^{0,1,2}_{gluex}$ we then inverted them as before into the corresponding six P-wave amplitudes, giving 10 free parameters in all with one phase fixed for each reflectivity. The results were as before; we get 2 complex conjugate solutions as shown in Figure \cref{fig:rho_sdme_t07}. After choosing one solution, they were then used to generate a new set of moments $H^{0,1,2,3}_{true}$, which, unlike the GlueX moments, exactly satisfy the constraint equations, since they are not subject to experimental noise.
In table \cref{tab:GXSDMEs} we give the original SDME values from the paper and the new values calculated from $H^{0,1,2,3}_{true}$. The recalculated values are all very close to the original data, all within one $\sigma$ of the given experimental uncertainties. This gives confidence that the underlying formalism, connecting partial waves to SDMEs and Spherical Harmonic moments, provides a self-consistent description of these data. In ~\cref{tab:pwa_values} we give the corresponding $P$-wave amplitude parameters.

\begin{table}[tbp]
    \centering
    \caption{Polarized $\rho$ SDMEs from GlueX at $t=-0.7278 \text{ GeV}^2$. The first row shows the actual GlueX results, while the second row show the values calculated from the amplitudes in \cref{tab:pwa_values}.}
     \label{tab:sdme_final_comparison}
    \begin{tabular}{|l|c|c|c|c|c|}
        \hline
        Source & $\rho^{0}_{00}$ & $\Re{\rho^{0}_{10}}$ & $\rho^{0}_{1-1}$ & $\rho^{1}_{00}$ & $\rho^{1}_{11}$ \\
        \hline
        Original Data & 0.0873 & 0.0257 & $-0.0211$ & $-0.0496$ & -0.0203 \\
        Uncert. Stat. $\pm$ &  0.0011& 0.0009& 0.0015& 0.0048 &0.0046 \\
        Uncert. Sys. $\pm$ &   0.0051 & 0.0014& 0.0015& 0.0037&0.0015 \\
        Recalculated & 0.0871 & 0.0259 & $-0.0198$ & $-0.0481$ & -0.0210 \\
        \hline \hline
        Source & $\rho^{1}_{10}$ & $\rho^{1}_{1-1}$ & $\Im{\rho^{2}_{10}}$ & $\Im{\rho^{2}_{1-1}}$ & \\
        \hline
        Original Data & $-0.0360$ & 0.4365 & 0.0179 & $-0.4119$ & \\
        Uncert. Stat. $\pm$ & 0.0029& 0.0074& 0.0034& 0.0052& \\
        Uncert. Syst. $\pm$ & 0.0015& 0.0094& 0.0015& 0.0088& \\
        Recalculated & $-0.0379$ & 0.4282 & 0.0146 & $-0.4185$ & \\
        \hline
    \end{tabular}
    \label{tab:GXSDMEs}
\end{table}

\begin{table}[tbp]
    \centering
    \caption{P waves extracted from GlueX P-waves SDME results at $t=-0.7278\text{ GeV}^2$. These values are used to generate data for subsequent tests.}
    \label{tab:pwa_values}
    \begin{tabular}{|l|c|c|}
        \hline
        Partial Wave & Magnitude & Phase (rad) \\
        \hline \hline
        $^{+}P_{-1}$ & 0.1033 & $-1.7829$ \\
        $^{+}P_{0}$  & 0.2606 & 1.4157 \\
        $^{+}P_{+1}$ & 0.9352 & 0.0000 \\
        \hline
        $^{-}P_{-1}$ & 0.0313 & 1.4686 \\
        $^{-}P_{0}$  & 0.1391 & $-2.3401$ \\
        $^{-}P_{+1}$ & 0.1627 & 0.0000 \\
        \hline
    \end{tabular}
\end{table}

\begin{table}[tbp]
    \centering
    \caption{Circularly Polarized $\rho$ SDMEs predicted from GlueX results at $t=-0.7278 \text{ GeV}^2$. The $\rho^3$ elements are calculated from the partial waves in \cref{tab:pwa_values}.}
    \begin{tabular}{|c|c|}
        \hline
        $\Im{\rho^{3}_{10}}$ & $\Im{\rho^{3}_{1-1}}$   \\
        \hline
         $-0.113$
          &   0.090 \\
        \hline
    \end{tabular}
    \label{tab:GXSDME3}
\end{table}

If we consider again the Intensity SDME equations ~\cref{eq:IrhoSDMEs} the $I^3(\Omega)$ dependence can be illustrated by forming a helicity asymmetry dependent on the circular polarization,
\begin{align} \nonumber
    \mathcal{A}_{3} &= \frac{\mathcal{I}(\Omega,P_L = 0, P_C=+1)-\mathcal{I}(\Omega,P_L = 0, P_C=-1)}{\mathcal{I}(\Omega,P_L = 0, P_C=+1)+\mathcal{I}(\Omega,P_L = 0,P_C=-1)} \\
    & = \frac{-\mathcal{I}^{3}(\Omega)}{\mathcal{I}_{0}(\Omega)}
\end{align}

\begin{figure}[tbp]
\begin{center}
\includegraphics[width=0.9\linewidth]{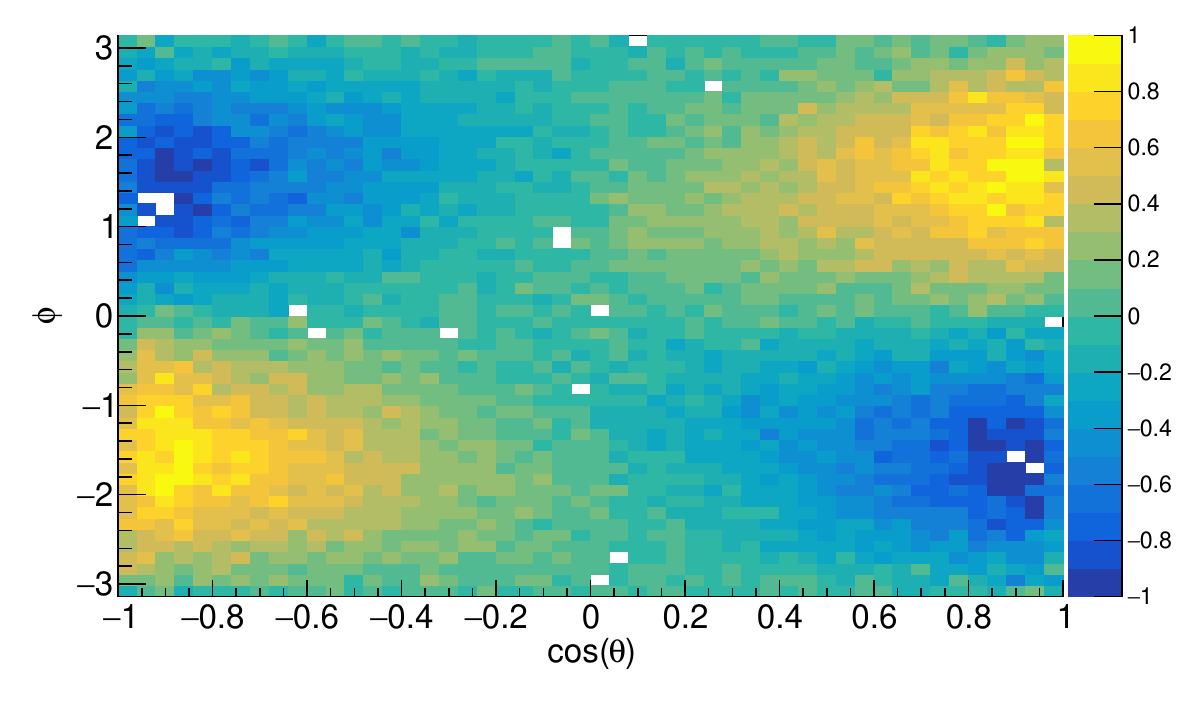}
\end{center}
\caption{The helicity asymmetry $\mathcal{A}_3(\Omega)$ predicted from the GlueX SDME results at $t=-0.7278 \text{ GeV}^2$. The distribution is characterised by the $\sin{2\theta}\sin{\phi}$ component related to a dominant $\Im{\rho^{3}_{10}}$.}\label{fig:A3} 
\end{figure}

A plot of the resulting 2D $\mathcal{A}_3(\Omega)$ distributions is given in \cref{fig:A3}. The dominant $\sin{2\theta}\sin{\phi}$ shape is due to the larger $\Im{\rho^{3}_{10}}$, as shown in \cref{tab:GXSDME3}, predicted from the GlueX results. The maximum and minimum values of this distribution are given by our analysis of the linearly polarized distribution. In an actual experiment, these values would be the product of the SDME distributions and the average degree of circular polarization in the data. Hence analysis of the decay distributions for elliptical polarizations gives a means to determine the degree of polarization. Next, we investigate this further.

\subsection{Amplitude and Polarizations extraction}

Considering the number of vector meson SDMEs we get nine experimentally measured quantities. This leads to ten constraint equations that relate moments to partial waves, as we have the further condition on the sum of the squares of the amplitude equaling the $H^0(0,0)$ moment. In each reflectivity, we have three partial waves from the three spin projections, yielding ten free parameters if we fix the phase of one partial wave in each reflectivity. So, these equations are just constrained and potentially invertible. We use the values for these given in ~\cref{tab:pwa_values}.

\subsubsection{Unknown Linear Polarization}
\begin{figure*}[tbp]
\centering
\includegraphics[width=\textwidth]{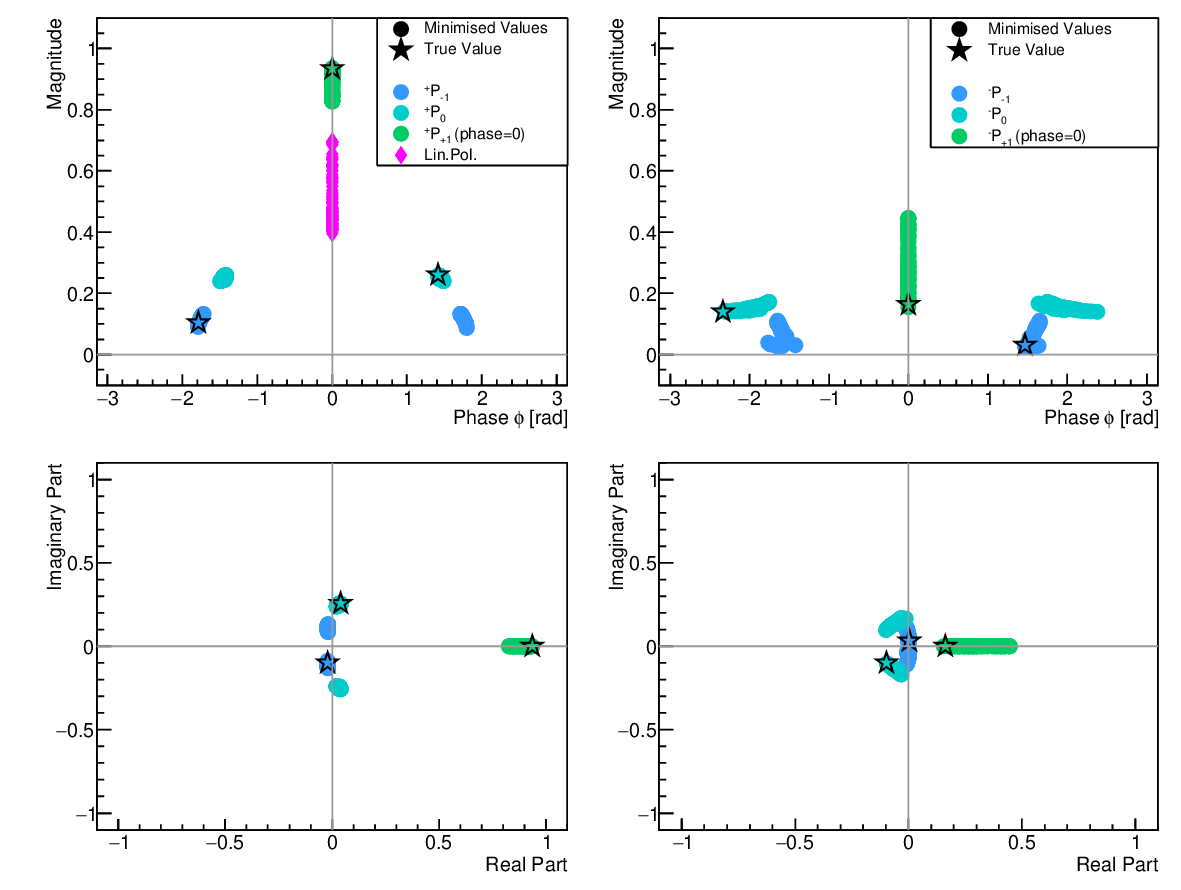}
\caption{
Resulting P-wave amplitudes from inverting moments calculated from the partial-waves in \cref{tab:pwa_values} with an additional unknown linear polarization.
The solid colored circles represent the results from one of the numerical inversions. The top row displays magnitude versus phase, and the bottom row shows the corresponding Argand diagrams. 
Left and right panels correspond to positive ($^{+}P_{\pm 1}$, etc.) and negative ($^{-}P_{\pm 1}$, etc.) reflectivity states, respectively.
}
\label{fig:rho_sdme_t07_plin}
\end{figure*}

In addition, we add a degree of linear polarization $P_{L}$ to the moments by multiplying $H^{1,2}(L,M)$ by that factor. For our tests, we choose $P_{L}=0.4$.
Now, when performing the inversion, we include an additional parameter to account for $P_{L}$, allowing us to extract the values for this consistent with the data. However, now with an additional parameter, our moment equations are no longer fully constrained and we do not expect to determine a precise value for $P_{L}$. The results are shown in \cref{fig:rho_sdme_t07_plin}. The unknown linear polarization gives a continuous ambiguity with a range of values allowed for all amplitude components. This is a similar effect to the continuous ambiguity shown in \cref{fig:pwa_h0h1h2_ambiguity} where the equations were also under-constrained.

\subsubsection{Unknown Linear and Circular Polarization}
\begin{figure*}[tbp]
\centering
\includegraphics[width=\textwidth]{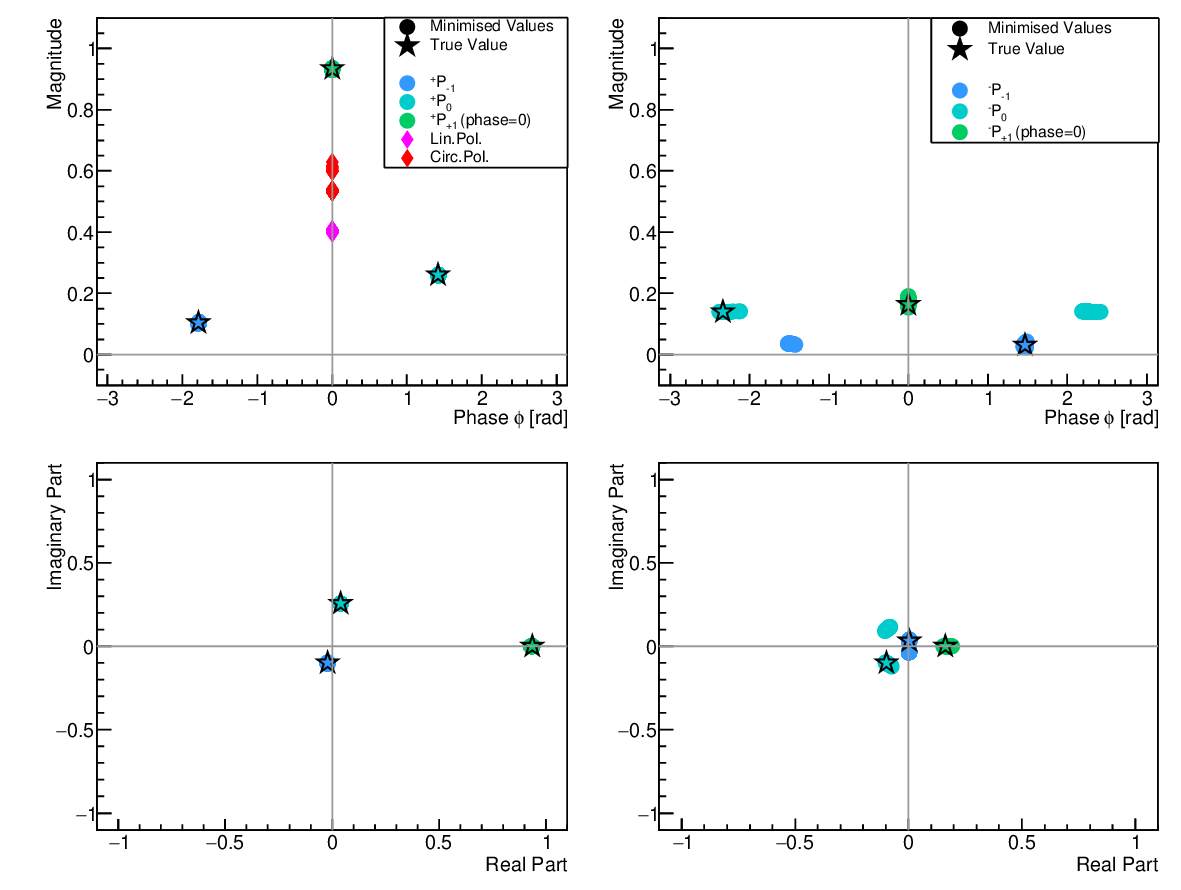}
\caption{
Resulting P-wave amplitudes from inverting moments calculated from the partial-waves in \cref{tab:pwa_values} with a additional unknown linear and circular polarizations.
The solid colored circles represent the results from one of the numerical inversions. The top row displays magnitude versus phase, and the bottom row shows the corresponding Argand diagrams. 
Left and right panels correspond to positive ($^{+}P_{\pm 1}$, etc.) and negative ($^{-}P_{\pm 1}$, etc.) reflectivity states, respectively.
}
\label{fig:rho_sdme_t07_plin_pcirc}
\end{figure*}

If we now also consider circular polarized moments, with an equivalent degree of polarization $P_{C}$, we have two further constraints, {$H^3(2,1)$, $H^3(2,1)$}, and one further unknown $P_{C}$. So now we have 12 unknowns and 12 constraints, the equations are now potentially solvable. If we numerically invert these we get the results shown in \cref{fig:rho_sdme_t07_plin_pcirc}. We see the positive reflectivity partial waves are well reproduced, in contrast the negative reflectivity partial waves have a resulting complex conjugate ambiguity, in addition there is a small range of allowed phases for these. The linear polarization is also well determined; however, we observe a discrete ambiguity in the circular polarization, which corresponds to either negative reflectivity complex conjugate solution.

In general, when performing measurements of amplitudes or SDMEs one has a number of kinematic bins for a given $W$ range, for example splitting in $t$ or invariant mass. In each bin, the underlying amplitudes should vary because of changes in the production mechanism or resonant state. By performing a similar analysis in multiple bins, the discrete ambiguity we have found may be resolved. 
As part of this study, the exercise was repeated, but this time with GlueX SDMEs in the bin $t=-0.23$~GeV$^2$. The results were similar with positive reflectivity amplitudes successfully determined and a complex conjugate solution for negative reflectivity. Here we just plot the values for the extracted circular and linear polarizations in both $t$ bins. The second results make clear that there are 2 solutions in the linear case as well, they just overlap in the first bin. It also shows that numerically there can be a wider range of similar solutions in different cases. In both cases, the true values for polarization were found, and by examining both, the true solution can be determined. Although it is not pursued here, fitting the partial waves with a mass or $t$ dependent model would naturally solve the ambiguities and allow the correct polarizations to be determined directly.

We also point out that the case of more partial waves, for example including S-waves along with $P$-waves, provides more constraints on the partial waves, beyond the case of just P-waves.~\cref{tab:counting} shows that for $S$- and $P$-waves we would have up to 18 measurable moments for 8 partial waves and 14 real numbers. However, difficulties may arise in some specific cases, for example, where some partial waves in the set are close to zero magnitude as discussed in ~\cref{Sec:pwavem=0}.

\begin{figure*}[tbp]
\centering
\includegraphics[width=\textwidth]{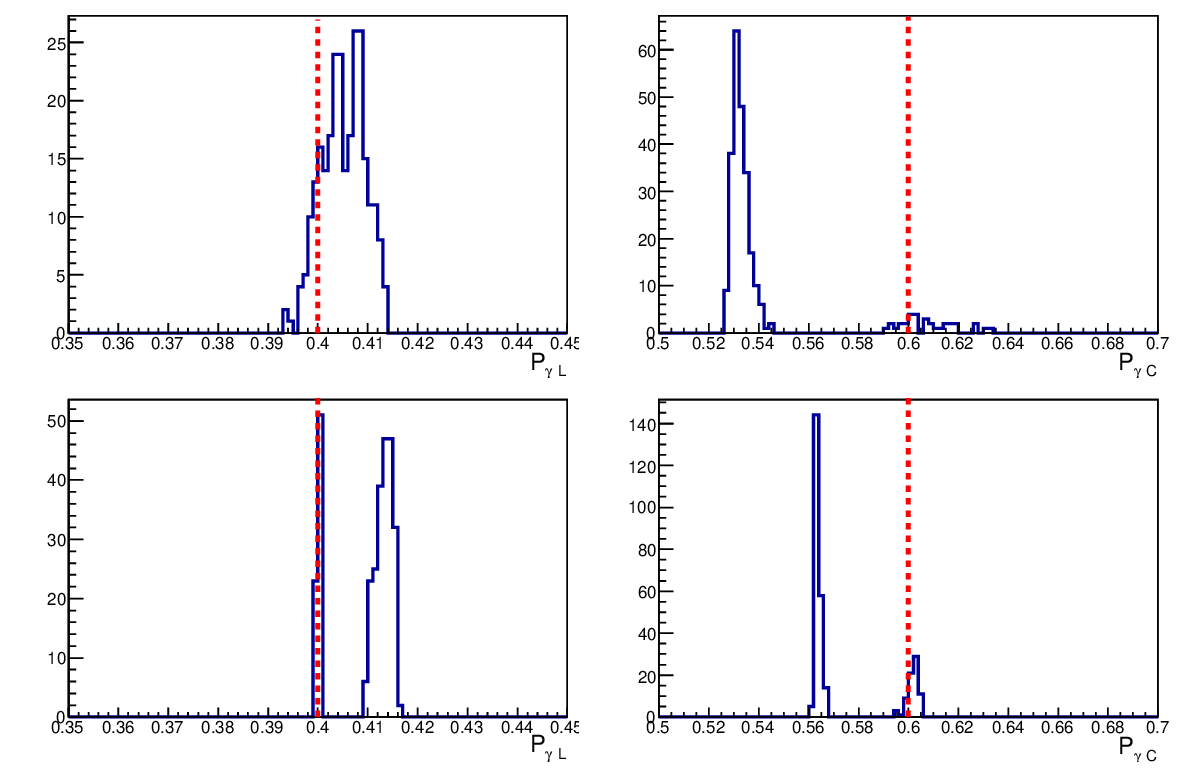}
\caption{
Extracted linear and circular degrees of polarization for studies with GlueX $\rho$ SDMEs for the $t=-0.728$ (top) and  $t=-0.230$ (bottom). Results of numerical inversion in blue. Dashed red line show true values.
}
\label{fig:plin_pcirc}
\end{figure*}

We note that determining external (nonamplitude) parameters, such as degrees of polarization, from fully constrained reactions is a methodology similar to that used in \cite{PhysRevLett.123.182301}. In this work $K\Lambda$ photoproduction was used to determine the $K\Lambda$ weak decay constant. There, the spin formalism for pseudoscalar meson production was used, and the previously measured polarization observables, akin to the SDMEs in the current analysis, which are dependent on underlying helicity amplitudes, provided sufficient constraints to allow determination of the decay constant.

\section{Summary}
The analysis of two-pseudoscalar photoproduction is complicated by the potential for mathematical ambiguities in the partial-wave expansion. This work followed on from recent studies that demonstrated how linearly polarized photon beams provide powerful constraints, yet can still permit ambiguities in certain scenarios~\cite{PhysRevD.108.076001, kjcp-h8b9}. By numerically inverting the equations that relate the experimental moments to the underlying partial-wave amplitudes, we have explicitly illustrated the nature of these ambiguities.

Our reanalysis of the $S$- and $D$-wave case in Section~\ref{sec:SDwaves} confirmed the findings of Smith et al.~\cite{PhysRevD.108.076001}: an unpolarized analysis is under-constrained, leading to a continuous ambiguity manifested as elliptical solution sets on the Argand diagram. The addition of full linear polarization resolves this continuous ambiguity but leaves a trivial complex conjugate ambiguity. We then examined the special case of P-waves from Guo et al.~\cite{kjcp-h8b9} in Section~\ref{Sec:pwavem=0} and again reproduced their finding of a continuous ambiguity even with linear polarization, a direct result of the system being under-constrained for that specific wave set.

The central contribution of this work is the inclusion of circular polarization. As derived in Appendix~\ref{app:h3_derivation}, the corresponding $H^3$ moments provide access to the imaginary parts of the amplitude bilinears, which are inaccessible with linear polarization alone. This additional information is precisely what is needed to break the remaining complex conjugate ambiguity. For the general case of $S$- and $D$-waves, the combination of all polarization observables creates an overconstrained system that yields a single, unique solution, removing the complex conjugate ambiguity entirely. In the special P-wave case, the addition of circular polarization resolves the continuous ambiguity into a two-fold discrete ambiguity, highlighting that while the system becomes fully determined, the nature of the solution space can still possess nontrivial features.

These findings have significant implications for future measurements at facilities like CLAS12, the EIC, and in future GlueX running, where both linear and circular polarization will be available. Just having additional constraints will reduce the uncertainty on amplitude parameters. The fact that the system becomes over-constrained opens up novel applications. As demonstrated in Section~\ref{Sec:pol_rho} using realistic amplitudes derived from GlueX $\rho$ photoproduction data~\cite{PhysRevC.108.055204}, it is possible to treat the degrees of linear and circular polarization as unknown parameters to be determined by the fit. This turns the reaction itself into a polarimeter, a technique that could be invaluable for precisely calibrating beam polarization. Although in this work we focussed on $\rho$ decaying to $\pi^+\pi^-$ it is likely that other reactions can provide more sensitivity due to differing amplitude combinations, for example a process with D-waves and higher negative reflectivity contributions may give clearer results. In any case, multiple reactions, as well as multiple kinmetic bins, may be used as a cross check for this procedure, giving strong constraints on systematic uncertainties.

Although the current study has focused on the mathematical framework, the additional constraints provided by elliptical polarization should also help to distinguish true solutions from false minima arising from experimental noise, an important consideration for real-world data analysis.

\acknowledgments
This material is based on work supported by the UK Science and Technology Facilities Council under grant ST/V00106X/1.
V.M. is a Serra Húnter Professor and acknowledges support from the Spanish national Grants PID2023-147112NB-C21 and CNS2022-136085.
We also acknowledge the use of Google's Gemini for assistance in structuring the appendices and refining the language of the manuscript. 

\appendix
\widetext
\section{The Elliptically Polarized Photon Spin-Density Matrix}
\label{app:EllipticalPhotonSDM}

\subsection{Degrees of Elliptical Polarization}
From the Stokes vectors for elliptical polarization the degrees of linear ($P_{L}$) and circular ($P_{C}$) polarization are related to the axis lengths $a$ and $b$ :
\begin{equation}
    P_{L} = \frac{a^2 - b^2}{a^2 + b^2}, \quad \quad P_{C} = \frac{2ab}{a^2 + b^2}.
\end{equation}
For a pure state where $P^{2}_{L} + P^{2}_{C} = 1$ and taking $a^2+b^2=1$, we can express the axis lengths in terms of the degree of linear polarization:
\begin{equation}\label{eq:ab_PcPl}
    a = \sqrt{\frac{1 + P_{L}}{2}} \quad \text{and} \quad b = \sqrt{\frac{1 - P_{L}}{2}}.
\end{equation}
And therefore in terms of $a$ alone we find,
\begin{equation}\label{eq:a_PcPl}
P_{L}=2a^2 -1 \quad \quad P_{C} = 2a\sqrt{1-a^2}
\end{equation}

\subsection{Connecting to the Schilling et al. Notation}
Using the photon density matrix from the formalism in Ref.~\cite{Schilling:1970vj}, the pure-state SDM is:
\begin{equation}
    \rho^{\text{pure}}(\gamma)=\frac{1}{2}\begin{pmatrix}
    1+2a\sqrt{1-a^{2}} & e^{-2i\Phi}(1-2a^{2}) \\
    e^{2i\Phi}(1-2a^{2}) & 1-2a\sqrt{1-a^{2}}
    \end{pmatrix}.
\end{equation}
Now using \cref{eq:a_PcPl} we can write the SDM in terms of experimentally measurable quantities $P_L$ and $P_C$:
\begin{equation}\label{eq:photo_rho_hel}
    \rho_{\lambda\lambda'} = \frac{1}{2} \begin{pmatrix}
    1 + P_{C} & -P_{L}e^{-i2\Phi} \\
    -P_{L}e^{i2\Phi} & 1 - P_{C}
    \end{pmatrix}.
\end{equation}

\section{The Quasi-Real Photon Limit for Virtual Photons}
\label{app:VirtualPhoton}

In electroproduction experiments, the exchanged photon is virtual. This appendix shows that in the limit of low momentum transfer ($Q^2 \to 0$), the virtual photon SDM becomes equivalent to the real photon SDM.

We start from the virtual photon density matrix as defined in Schilling and Wolf~\cite{Schilling:1970vj}. In the helicity basis ($\lambda=1,0,-1$), this is given by:
\begin{equation}
    \rho(\gamma)_{\lambda\lambda'} = \frac{1}{2}
    \begin{pmatrix}
    1+\sqrt{1-\epsilon^{2}}P_{e} &
    \sqrt{\epsilon}e^{-i\Phi} \left( \sqrt{1+\epsilon} + \sqrt{1-\epsilon}P_{e} \right) &
    -\epsilon e^{-2i\Phi}  \\
    \sqrt{\epsilon}e^{i\Phi} \left( \sqrt{1+\epsilon} + \sqrt{1-\epsilon}P_{e} \right)&
    2\epsilon &
    -\sqrt{\epsilon(1+\epsilon)}e^{-i\Phi} + \sqrt{\epsilon}e^{i\Phi}\sqrt{1-\epsilon}P_{e}
    \\
    -\epsilon e^{2i\Phi}  &
    -\sqrt{\epsilon(1+\epsilon)}e^{i\Phi} + \sqrt{\epsilon}e^{-i\Phi}\sqrt{1-\epsilon}P_{e} &
    1-\sqrt{1-\epsilon^{2}}P_{e}
    \end{pmatrix}
\end{equation}
where $\epsilon$ is the transverse polarization parameter of the virtual photon and $P_e$ is the polarization of the electron beam.

In the case of quasi-real photoproduction ($Q^2 \to 0$), the contributions from the longitudinal photon component ($\lambda=0$) are suppressed. Removing the rows and columns corresponding to $\lambda=0$ leaves the $2 \times 2$ matrix for the transverse components:
\begin{equation}
    \rho(\gamma)_{\lambda\lambda'} = \frac{1}{2}
    \begin{pmatrix}
    1+\sqrt{1-\epsilon^{2}}P_{e} &
    -\epsilon e^{-2i\Phi}  \\
    -\epsilon e^{2i\Phi}  &
     1-\sqrt{1-\epsilon^{2}}P_{e}
    \end{pmatrix}.
\end{equation}
By comparing this to the real photon SDM in Eq.~\ref{eq:photo_rho_hel}, we can identify the equivalent degrees of polarization for the quasi-real photon:
\begin{itemize}
    \item Degree of Linear Polarization: $P_{L} = \epsilon$
    \item Degree of Circular Polarization: $P_{C} = \sqrt{1-\epsilon^{2}}P_{e}$
\end{itemize}
This demonstrates that analyses of electroproduction data in the quasi-real limit can be treated with the same formalism developed for real photoproduction.

\section{Derivation of the H\textsuperscript{3} Moment Expression}
\label{app:h3_derivation}
Here we prove that the circularly polarized moment, $H^3(L,M)$, is purely imaginary by showing that it can be expressed as a difference of a term and its complex conjugate. This follows the similar derivation in Guo et. al. \cite{kjcp-h8b9} which proves the other moments are all real.

The derivation begins with the expression for $H^3(L,M)$ for a single reflectivity component $\epsilon$, which is constructed from the spin density matrix element $\rho^3$ \cite{Mathieu:2019fts}. Constants of proportionality have been omitted for the sake of clarity.
\begin{equation}
    H^3(L,M) \propto - \sum_{\ell,m,\ell',m'} C_{\ell m;\ell'm'}^{LM} \left( [\ell]_{m}^{(\epsilon)}[\ell']_{m'}^{(\epsilon)*} - (-1)^{m-m'}[\ell]_{-m}^{(\epsilon)}[\ell']_{-m'}^{*} \right)
\end{equation}

We can decompose this into two parts:
\begin{equation}
    H^3(L,M) \propto - \left[ \left( \sum_{\ell,m,\ell',m'} C_{\ell m;\ell'm'}^{LM} [\ell]_{m}^{(\epsilon)}[\ell']_{m'}^{(\epsilon)*} \right) - \left( \sum_{\ell,m,\ell',m'} C_{\ell m;\ell'm'}^{LM} (-1)^{m-m'}[\ell]_{-m}^{(\epsilon)}[\ell']_{-m'}^{*} \right) \right]
\end{equation}

The key step in Guo et al. involves manipulating the second summation term by relabeling the dummy summation indices and applying the symmetry properties of the Clebsch-Gordan coefficients; the second term is shown to be the complex conjugate of the first term.

Taking just the second term, we relabel the dummy summation indices and apply the fundamental symmetry property of the Clebsch-Gordan coefficients:
\begin{equation*}
    \langle j_1 m_1, j_2 m_2 | J M \rangle = (-1)^{j_2+m_2} \sqrt{\frac{2J+1}{2j_1+1}} \langle J(-M), j_2 m_2 | j_1(-m_1) \rangle
\end{equation*}

This manipulation transforms the coefficient back to its original form, $C_{lm;l'm'}^{LM}$, but introduces a phase factor $(-1)^{2L+M}$. The entire sum becomes:
\begin{equation*}
    \sum_{\ell,m,\ell',m'} (-1)^{2L+M}C_{\ell m;\ell'm'}^{LM} (-1)^{-m+m'}[\ell']_{m'}^{(\epsilon)}[\ell]_{m}^{(\epsilon)*}
\end{equation*}

The Clebsch-Gordan coefficients are non-zero only if angular momentum is conserved, which enforces the selection rule $M = m - m'$. We isolate the phase factors and substitute this constraint in:
\begin{equation*}
    (-1)^{2L+M} (-1)^{-m+m'} = (-1)^{2L+(m-m')} (-1)^{-(m-m')} = (-1)^{2L} = +1
\end{equation*}
Hence, the phase factors cancel completely.

The second term then becomes the exact complex conjugate of the first term:
\begin{equation} \label{eq:conjugate_identity}
    \sum_{\ell,m,\ell',m'} C_{\ell m;\ell 'm'}^{LM} (-1)^{m-m'}[\ell]_{-m}^{(\epsilon)}[\ell']_{-m'}^{*} = \left( \sum_{\ell,m,\ell',m'} C_{\ell m;\ell 'm'}^{LM} [\ell]_{m}^{(\epsilon)}[\ell']_{m'}^{(\epsilon)*} \right)^*
\end{equation}

Substituting this identity (\ref{eq:conjugate_identity}) back into the expression for $H^3(L,M)$, we obtain a result in the form of a term minus its complex conjugate. If we define a complex term $Z$ as:
\begin{equation}
    Z = \sum_{\ell,m,\ell',m'} C_{\ell m;\ell'm'}^{LM} [\ell]_{m}^{(\epsilon)}[\ell']_{m'}^{(\epsilon)*}
\end{equation}
Then the expression for the moment becomes:
\begin{equation}
    H^3(L,M) \propto -(Z - Z^*)
\end{equation}
Since $Z - Z^* = 2i \cdot \text{Im}(Z)$, the final expression is:
\begin{equation}
    H^3(L,M) \propto -2i \cdot \text{Im}(Z)
\end{equation}
This proves that the $H^3(L,M)$ moment is purely imaginary and related to the imaginary parts of the partial wave billinear terms.

This result is analogous to the expansion of the real moment $H_0$, which is constructed from a linear combination of the \textbf{real parts} of the amplitude products. For the purely imaginary moment $H^3$, the expansion is instead constructed from a linear combination of the \textbf{imaginary parts} of the same products. The corresponding real-valued building blocks for this expansion are:
\begin{equation}
    \im\left([\ell]_{m}^{(\epsilon)}[\ell^{\prime}]_{m^{\prime}}^{(\epsilon)*}\right) = 
    \begin{cases} 
        0 & \text{if } l=l' \text{ and } m=m' \\ |A_{\ell,m}^{(\epsilon)}||A_{\ell',m'}^{(\epsilon)}|\sin(\phi_{\ell,m}^{(\epsilon)}-\phi_{\ell',m'}^{(\epsilon)}) & \text{if } \ell\neq \ell' \text{ or } m\neq m' 
    \end{cases}
\end{equation}
where $|A|$ and $\phi$ are the magnitude and phase of the partial-wave amplitude, respectively. This highlights that the imaginary nature of $H^3$ is fundamentally linked to the sine of the relative phases between interfering partial waves.

\section{Solving for Amplitudes in the S- and D-Wave Model}
\label{app:sd_wave_solution}

This appendix provides a step-by-step analytical procedure for determining the amplitude parameters for the wave set $\{S_0^+, D_{+1}^+, D_0^+, D_{-1}^+\}$ from the measured moments. As shown in Ref.~\cite{kjcp-h8b9}, this system is overconstrained when all polarization observables are included, meaning there is no continuous mathematical ambiguity. The following procedure demonstrates how to uniquely determine all amplitude magnitudes and relative phases. We begin by fixing the overall phase of the S-wave, $\phi_{S_0}^{+}$, to zero without loss of generality.

The non-zero moments for this wave set are given by:
\subsection*{$L=0$ Moments}
\begin{align*}
    H^{0}(0,0) &= 2 \left( |S_{0}^{+}|^2 + |D_{+1}^{+}|^2 + |D_{0}^{+}|^2 + |D_{-1}^{+}|^2 \right) \\
    H^{1}(0,0) &= 2 \left( |S_{0}^{+}|^2 + |D_{0}^{+}|^2 - 2 |D_{+1}^{+}||D_{-1}^{+}|\cos(\phi_{D_{+1}}^{+} - \phi_{D_{-1}}^{+}) \right)
\end{align*}

\subsection*{$L=2$ Moments}
\begin{align*}
    H^{0}(2,0) &= 2 \left( \frac{2\sqrt{5}}{5}|S_{0}^{+}||D_{0}^{+}|\cos(\phi_{D_0}^{+}) + \frac{2}{7}|D_{0}^{+}|^2 + \frac{1}{7}|D_{+1}^{+}|^2 + \frac{1}{7}|D_{-1}^{+}|^2 \right) \\
    H^{1}(2,0) &= 2 \left( \frac{2\sqrt{5}}{5}|S_{0}^{+}||D_{0}^{+}|\cos(\phi_{D_0}^{+}) + \frac{1}{7}|D_{0}^{+}|^2 - \frac{1}{7}|D_{+1}^{+}||D_{-1}^{+}|\cos(\phi_{D_{+1}}^{+} - \phi_{D_{-1}}^{+}) \right) \\
    H^{0}(2,1) &= 2 \left( \frac{\sqrt{5}}{5}\left(|S_{0}^{+}||D_{+1}^{+}|\cos(\phi_{D_{+1}}^{+}) - |S_{0}^{+}||D_{-1}^{+}|\cos(\phi_{D_{-1}}^{+})\right) \right. \\
    & \quad \left. + \frac{1}{7}\left(|D_{+1}^{+}||D_{0}^{+}|\cos(\phi_{D_{+1}}^{+} - \phi_{D_0}^{+}) - |D_{0}^{+}||D_{-1}^{+}|\cos(\phi_{D_0}^{+} - \phi_{D_{-1}}^{+})\right) \right) \\
    H^{1}(2,1) &= 2 \left( \frac{\sqrt{5}}{5}\left(|S_{0}^{+}||D_{+1}^{+}|\cos(\phi_{D_{+1}}^{+}) - |S_{0}^{+}||D_{-1}^{+}|\cos(\phi_{D_{-1}}^{+})\right) \right. \\
    & \quad \left. - \frac{1}{7}\left(|D_{+1}^{+}||D_{0}^{+}|\cos(\phi_{D_{+1}}^{+} - \phi_{D_0}^{+}) - |D_{0}^{+}||D_{-1}^{+}|\cos(\phi_{D_0}^{+} - \phi_{D_{-1}}^{+})\right) \right) \\
    \frac{H^{2}(2,1)}{i} &= 2 \left( -\frac{\sqrt{5}}{5}\left(|S_{0}^{+}||D_{+1}^{+}|\cos(\phi_{D_{+1}}^{+}) + |S_{0}^{+}||D_{-1}^{+}|\cos(\phi_{D_{-1}}^{+})\right) \right. \\
    & \quad \left. - \frac{1}{7}\left(|D_{+1}^{+}||D_{0}^{+}|\cos(\phi_{D_{+1}}^{+} - \phi_{D_0}^{+}) + |D_{0}^{+}||D_{-1}^{+}|\cos(\phi_{D_0}^{+} - \phi_{D_{-1}}^{+})\right) \right) \\
    \frac{H^{3}(2,1)}{i} &= 2 \left( \frac{\sqrt{5}}{5}\left(|S_{0}^{+}||D_{+1}^{+}|\sin(\phi_{D_{+1}}^{+}) - |S_{0}^{+}||D_{-1}^{+}|\sin(\phi_{D_{-1}}^{+})\right) \right. \\
    & \quad \left. + \frac{1}{7}\left(|D_{+1}^{+}||D_{0}^{+}|\sin(\phi_{D_{+1}}^{+} - \phi_{D_0}^{+}) - |D_{0}^{+}||D_{-1}^{+}|\sin(\phi_{D_0}^{+} - \phi_{D_{-1}}^{+}) \right) \right)
\end{align*}

\subsection*{$L=4$ Moments}
\begin{align*}
    H^{1}(4,1) &= \frac{2\sqrt{30}}{21} \left(-|D_{+1}^{+}||D_{0}^{+}|\cos(\phi_{D_{+1}}^{+} - \phi_{D_0}^{+}) + |D_{0}^{+}||D_{-1}^{+}|\cos(\phi_{D_0}^{+} - \phi_{D_{-1}}^{+})\right) \\
    \frac{H^{2}(4,1)}{i} &= \frac{2\sqrt{30}}{21} \left(-|D_{+1}^{+}||D_{0}^{+}|\cos(\phi_{D_{+1}}^{+} - \phi_{D_0}^{+}) - |D_{0}^{+}||D_{-1}^{+}|\cos(\phi_{D_0}^{+} - \phi_{D_{-1}}^{+})\right) \\
    \frac{H^{3}(4,1)}{i} &= \frac{2\sqrt{30}}{21} \left(|D_{+1}^{+}||D_{0}^{+}|\sin(\phi_{D_{+1}}^{+} - \phi_{D_0}^{+}) - |D_{0}^{+}||D_{-1}^{+}|\sin(\phi_{D_0}^{+} - \phi_{D_{-1}}^{+})\right) \\
    H^{0}(4,2) &= -\frac{4\sqrt{10}}{21}|D_{+1}^{+}||D_{-1}^{+}|\cos(\phi_{D_{+1}}^{+} - \phi_{D_{-1}}^{+}) \\
    H^{1}(4,2) &= \frac{2\sqrt{10}}{21} \left( \sqrt{2}|D_{+1}^{+}|^2 + \sqrt{2}|D_{-1}^{+}|^2 \right) \\
    \frac{H^{2}(4,2)}{i} &= \frac{2\sqrt{10}}{21} \left( -\sqrt{2}|D_{+1}^{+}|^2 + \sqrt{2}|D_{-1}^{+}|^2 \right) \\
    \frac{H^{3}(4,2)}{i} &= -\frac{4\sqrt{10}}{21}|D_{+1}^{+}||D_{-1}^{+}|\sin(\phi_{D_{+1}}^{+} - \phi_{D_{-1}}^{+})
\end{align*}
\begin{enumerate}
    \item \textbf{Determine the magnitudes of the $\boldsymbol{m=\pm 1}$ D-waves:}
    The magnitudes of $|D_{+1}^{+}|$ and $|D_{-1}^{+}|$ can be isolated by forming a simple linear combination of the $L=2, M=2$ moments. The equations for $H^1(2,2)$ and $H^2(2,2)$ form a system of two linear equations for the two unknowns $|D_{+1}^{+}|^2$ and $|D_{-1}^{+}|^2$.
    \begin{align}
        |D_{+1}^{+}|^2 &= \frac{7}{2\sqrt{6}} \left( H^{1}(2,2) - \frac{H^{2}(2,2)}{i} \right) \\
        |D_{-1}^{+}|^2 &= \frac{7}{2\sqrt{6}} \left( H^{1}(2,2) + \frac{H^{2}(2,2)}{i} \right)
    \end{align}

    \item \textbf{Determine the magnitude of the $\boldsymbol{m=0}$ D-wave:}
    With the magnitudes of the $m=\pm 1$ D-waves known, the moment $H^0(4,0)$ can be used to solve for the magnitude of the $D_0^+$ wave.
    \begin{equation}
        |D_{0}^{+}|^2 = \frac{1}{6} \left( \frac{21 H^{0}(4,0)}{2} + 4|D_{+1}^{+}|^2 + 4|D_{-1}^{+}|^2 \right)
    \end{equation}

    \item \textbf{Determine the magnitude of the S-wave:}
    With all three D-wave magnitudes determined, the total intensity moment, $H^0(0,0)$, is used to solve for the S-wave magnitude.
    \begin{equation}
        |S_{0}^{+}|^2 = \frac{H^{0}(0,0)}{2} - \left( |D_{+1}^{+}|^2 + |D_{0}^{+}|^2 + |D_{-1}^{+}|^2 \right)
    \end{equation}

    \item \textbf{Determine the relative phases:}
    At this point, all four magnitudes are known. The remaining moments, which contain interference terms, can be used to solve for the relative phases. Using $H^0$ and $H^3$ moments provides constraints on both the cosine and sine of the phase differences, respectively. For example, for the phase difference $\Delta\phi = (\phi_{D_{+1}}^{+} - \phi_{D_{-1}}^{+})$:
    \begin{align}
        \cos(\Delta\phi) &= -\frac{7 H^{0}(2,2)}{2\sqrt{6}|D_{+1}^{+}||D_{-1}^{+}|} \\
        \sin(\Delta\phi) &= -\frac{7 H^{3}(2,2)/i}{2\sqrt{6}|D_{+1}^{+}||D_{-1}^{+}|}
    \end{align}
    Knowing both the sine and cosine of an angle uniquely determines the angle in the range $[-\pi, \pi]$. The same principle applies to all other phase differences using the other interference moments. The unique phase difference is given by:
    \begin{equation}
        \Delta\phi = \tan^{-1}\left(\frac{\sin(\Delta\phi)}{\cos(\Delta\phi)}\right)
    \end{equation}
    This procedure removes the sign ambiguity for each phase, resolving the complex conjugate ambiguity and leading to a single, unique solution for all amplitude parameters.
\end{enumerate}
\section{Constraints on Amplitudes in the Simplified P-Wave Model}
\label{app:p_wave_constraints}

This appendix details the derivation of constraints on partial-wave amplitudes in the presence of the continuous mathematical ambiguity found in the simplified wave set $\{P_{+1}^{\pm}, P_{-1}^{\pm}\}$.

\subsection{Simplified Moment Equations}
For this wave set, the five independent, non-zero moments are:
\begin{align*}
    H^{0}(0,0) &= 2 \left( |P_{+1}^{+}|^2 + |P_{-1}^{+}|^2 \right) + 2 \left( |P_{+1}^{-}|^2 + |P_{-1}^{-}|^2 \right) \\
    H^{1}(2,2) &= \frac{\sqrt{6}}{5} \left( |P_{+1}^{+}|^2 + |P_{-1}^{+}|^2 \right) - \frac{\sqrt{6}}{5} \left( |P_{+1}^{-}|^2 + |P_{-1}^{-}|^2 \right) \\
    H^{2}(2,2)/i &= \frac{\sqrt{6}}{5} \left( |P_{-1}^{+}|^2 - |P_{+1}^{+}|^2 \right) - \frac{\sqrt{6}}{5} \left( |P_{-1}^{-}|^2 - |P_{+1}^{-}|^2 \right) \\
    H^{0}(2,2) &= -\frac{2\sqrt{6}}{5} \left( |P_{+1}^{+}||P_{-1}^{+}|\cos(\phi_{P_{-1}}^{+}) + |P_{+1}^{-}||P_{-1}^{-}|\cos(\phi_{P_{-1}}^{-}) \right) \\
    H^{1}(0,0) &= -4 |P_{+1}^{+}||P_{-1}^{+}|\cos(\phi_{P_{-1}}^{+}) + 4 |P_{+1}^{-}||P_{-1}^{-}|\cos(\phi_{P_{-1}}^{-})
\end{align*}
This system has fewer independent equations (five) than unknown parameters (six: four magnitudes and two relative phases), leading to a continuous ambiguity. We can, however, derive the exact ranges that bound the solutions for the amplitude parameters.

\subsection{Derivation of Amplitude Ranges}

\begin{enumerate}
    \item \textbf{Define Physical Observables for Each Reflectivity:}
    We first define physical observables for each reflectivity sector ($\epsilon = +, -$). These are the sum of intensities ($I^{\epsilon}$), the difference of intensities ($M^{\epsilon}$), and the real part of the interference term ($R^{\epsilon}$):
    \begin{align*}
        I^{\epsilon} &= |P_{+1}^{\epsilon}|^2 + |P_{-1}^{\epsilon}|^2 \\
        M^{\epsilon} &= |P_{-1}^{\epsilon}|^2 - |P_{+1}^{\epsilon}|^2 \\
        R^{\epsilon} &= |P_{+1}^{\epsilon}||P_{-1}^{\epsilon}|\cos(\phi_{P_{-1}}^{\epsilon})
    \end{align*}

    \item \textbf{Solve for the Observable Combinations:}
    By forming linear combinations of the moment equations, we can solve for five combinations of these observables:
    \begin{align}
        I^{+} + I^{-} &= \frac{1}{2}H_{0}(0,0) \\
        I^{+} - I^{-} &= \frac{5}{\sqrt{6}}H_{1}(2,2) \\
        R^{+} &= -\frac{1}{8} \left( \frac{10}{\sqrt{6}}H_{0}(2,2) + H_{1}(0,0) \right) \\
        R^{-} &= \frac{1}{8}H_{1}(0,0) - \frac{5}{4\sqrt{6}}H_{0}(2,2) \\
        M^{+} - M^{-} &= \frac{5}{i\sqrt{6}}H_{2}(2,2)
    \end{align}
    From the first two equations, we can determine $I^{+}$ and $I^{-}$ individually. However, we can only solve for the \textit{difference} $M^{+} - M^{-}$, which is the source of the continuous ambiguity.

    \item \textbf{Establish Reflectivity-Independent Constraints:}
    The condition $|\cos(\phi_{P_{-1}}^{\epsilon})| \le 1$ implies that $(R^{\epsilon})^2 \le (|P_{+1}^{\epsilon}||P_{-1}^{\epsilon}|)^2$. Using the identity $(I^{\epsilon})^2 - (M^{\epsilon})^2 = 4(|P_{+1}^{\epsilon}||P_{-1}^{\epsilon}|)^2$, we find the allowed range for $M^{\epsilon}$:
    \[ -\sqrt{(I^{\epsilon})^2 - 4(R^{\epsilon})^2} \le M^{\epsilon} \le \sqrt{(I^{\epsilon})^2 - 4(R^{\epsilon})^2} \]
    This, in turn, provides a symmetric allowed range for the squared magnitudes:
    \begin{equation}\label{eq:range1_appE}
        \frac{I^{\epsilon} - \sqrt{(I^{\epsilon})^2 - 4(R^{\epsilon})^2}}{2} \le |P_{+1}^{\epsilon}|^2, |P_{-1}^{\epsilon}|^2 \le \frac{I^{\epsilon} + \sqrt{(I^{\epsilon})^2 - 4(R^{\epsilon})^2}}{2}
    \end{equation}

    \item \textbf{Link Reflectivity Sectors and Determine Final Ranges:}
    The ambiguity is coupled via the linking constraint $M^{+} = (M^{+} - M^{-}) + M^{-}$. Using the allowed range for $M^-$ (from Step 3) and the known value of $(M^{+} - M^{-})$ (from Step 2), we can determine the required range for $M^{+}$. A solution is only physical if $|M^{+}| \le I^{+}$. The final, valid range for $M^{+}$ is the intersection of the required and physical ranges:
    \[ M^{+}_{\text{final\_range}} = [M^{+}_{\min, \text{req}}, M^{+}_{\max, \text{req}}] \cap [-I^{+}, +I^{+}] \]
    Using this final range for $M^{+}$, we can find the linked-constraint ranges for the squared magnitudes:
    \begin{itemize}
        \item Range for $|P_{+1}^{+}|^2$: $[\frac{1}{2}(I^{+} - M^{+}_{\max}), \frac{1}{2}(I^{+} - M^{+}_{\min})]$
        \item Range for $|P_{-1}^{+}|^2$: $[\frac{1}{2}(I^{+} + M^{+}_{\min}), \frac{1}{2}(I^{+} + M^{+}_{\max})]$
    \end{itemize}
    The most constrained, true allowed range for each magnitude is the \textbf{intersection} of this new range with the one found in Step 3. This procedure is then repeated for the negative-reflectivity amplitudes.
\end{enumerate}

\subsection{Numerical Example}
We use the input amplitude values from Ref.~\cite{kjcp-h8b9}. With the constraint $\sum |A|^2 = 1$, we have $|P_{+1}^{+}| \approx 0.8598$.
\begin{itemize}
    \item $|P_{-1}^{+}| = 0.130$
    \item $\phi_{P_{-1}}^{+} = -1.1135$ rad
    \item $|P_{+1}^{-}| = 0.142$
    \item $|P_{-1}^{-}| = 0.473$
    \item $\phi_{P_{-1}}^{-} = -1.6354$ rad
\end{itemize}
These amplitudes correspond to the following numerical values for the observables:
\begin{itemize}
    \item $I^{+} \approx 0.7561$
    \item $I^{-} \approx 0.2439$
    \item $R^{+} \approx 0.0493$
    \item $R^{-} \approx -0.0043$
    \item $M^{+} - M^{-} \approx -0.9259$
\end{itemize}
The final intersected ranges derived from these values are found to be:
\begin{itemize}
    \item \textbf{Range($|P_{+1}^{+}|$):} \textbf{[0.8480, 0.8677]}
    \item \textbf{Range($|P_{-1}^{+}|$):} \textbf{[0.0569, 0.1923]}
    \item \textbf{Range($|P_{+1}^{-}|$):} \textbf{[0.0088, 0.1839]}
    \item \textbf{Range($|P_{-1}^{-}|$):} \textbf{[0.4583, 0.4938]}
\end{itemize}
And the corresponding phase ranges are:
\begin{itemize}
    \item \textbf{For $\phi_{P_{-1}}^{+}$:} The value of $\cos(\phi_{P_{-1}}^{+})$ is constrained to \textbf{[0.302, 1.0]}. This corresponds to a phase angle range of \textbf{[-1.263, +1.263] radians} or \textbf{[-72.4$^{\circ}$, +72.4$^{\circ}$]}.
    \item \textbf{For $\phi_{P_{-1}}^{-}$:} The value of $\cos(\phi_{P_{-1}}^{-})$ is constrained to \textbf{[-1.0, -0.051]}. This corresponds to a phase angle range of \textbf{[-3.142, -1.622] $\cup$ [1.622, 3.142] radians} or \textbf{[-180.0$^{\circ}$, -92.9$^{\circ}$] $\cup$ [92.9$^{\circ}$, 180.0$^{\circ}$]}.
\end{itemize}

\section{Resolving the Continuous Ambiguity with Circular Polarization}
\label{app:resolving_ambiguity}

This appendix demonstrates how the inclusion of circular polarization resolves the continuous ambiguity described in Appendix~\ref{app:p_wave_constraints}, reducing it to a two-fold discrete ambiguity.

\subsection{The Additional Constraint}
The introduction of circular polarization provides one additional non-zero moment, $H^3(2,2)$, which constrains the imaginary part of the interference terms. We define a new physical observable, $S^{\epsilon}$:
\[ S^{\epsilon} = |P_{+1}^{\epsilon}||P_{-1}^{\epsilon}|\sin(\phi_{P_{+1}}^{\epsilon} - \phi_{P_{-1}}^{\epsilon}) \]
From the expression for $H^3(2,2)/i$ in Appendix~\ref{App:Spin1H}, we can solve for the sum of these terms:
\[ S^{+} + S^{-} = -\frac{5}{2\sqrt{6}} \frac{H^{3}(2,2)}{i} \]

\subsection{Solving the System of Equations}
To find the discrete solutions, first we derive the following four identities which we must solve as a complete system of equations for the four unknown quantities $M^+, M^-, S^+, S^-$:
\begin{align}
    M^+ - M^- &= C_1 \label{eq:res_1} \\
    S^+ + S^- &= C_2 \label{eq:res_2} \\
    (M^+)^2 + 4((S^+)^2 + (R^+)^2) &= (I^+)^2 \label{eq:res_3} \\
    (M^-)^2 + 4((S^-)^2 + (R^-)^2) &= (I^-)^2 \label{eq:res_4}
\end{align}
where $C_1$ and $C_2$ are the known values determined from the $H^2(2,2)$ and $H^3(2,2)$ moments, respectively. The third and fourth equations are derived from an identity relating the sum of intensities ($I^\epsilon$), the difference of intensities ($M^\epsilon$), and the product of the magnitudes. This identity is derived by noting that $(I^\epsilon)^2 = (|P_{+1}^{\epsilon}|^2 + |P_{-1}^{\epsilon}|^2)^2$ and $(M^\epsilon)^2 = (|P_{-1}^{\epsilon}|^2 - |P_{+1}^{\epsilon}|^2)^2$, which can be rearranged to show:
\[ (M^\epsilon)^2 + 4(|P_{+1}^{\epsilon}||P_{-1}^{\epsilon}|)^2 = (I^\epsilon)^2 \]
Combining this with $(|P_{+1}^{\epsilon}||P_{-1}^{\epsilon}|)^2 = (R^{\epsilon})^2 + (S^{\epsilon})^2$ , which uses $\cos^2{\phi}+\sin^2{\phi} = 1$, yields Eqs.~\eqref{eq:res_3} and \eqref{eq:res_4}.

We can express $M^-$ and $S^-$ in terms of $M^+$ and $S^+$ using Eqs.~\ref{eq:res_1} and \ref{eq:res_2}, and substitute them into Eq.~\eqref{eq:res_4}:
\[ (M^+ - C_1)^2 + 4((C_2 - S^+)^2 + (R^-)^2) = (I^-)^2 \]
This equation, along with Eq.~\eqref{eq:res_3}, forms a system of two quadratic equations for the two variables $M^+$ and $S^+$. Geometrically, this is the intersection of two ellipses. We can solve this system by substitution. From Eq.~\eqref{eq:res_3}, we can express $(M^+)^2$ in terms of $(S^+)^2$. Substituting this into the expanded form of the equation above yields a quadratic equation for the variable $S^+$. The two roots of this quadratic equation provide the two possible values for $S^+$.

For each of the two solutions for $S^+$, a corresponding value for $M^+$ is found. Subsequently, the values for $M^-$ and $S^-$ are determined. Finally, the magnitudes and phases for each of the two discrete solutions can be calculated using the relations:
\begin{align*}
    |P_{+1}^{\epsilon}|^2 &= \frac{1}{2}(I^{\epsilon} - M^{\epsilon}) & |P_{-1}^{\epsilon}|^2 &= \frac{1}{2}(I^{\epsilon} + M^{\epsilon}) &
    \phi_{P_{-1}}^{\epsilon} &= \arctan(S^{\epsilon}, R^{\epsilon})
\end{align*}

\subsection{Numerical Example of the Two-Fold Ambiguity}
Using the same initial values from the numerical example in Appendix~\ref{app:p_wave_constraints}, the continuous range of possibilities collapses to the two distinct, discrete solutions presented below.

\begin{table}[h!]
\centering
\caption{Solution 1, corresponding to the initial input values.}
\begin{tabular}{|c|c|c|c|}
\hline
\textbf{Amplitude} & \textbf{Magnitude} & \textbf{Phase (radians)} & \textbf{Phase (degrees)} \\
\hline
$|P_{+1}^{+}|$ & 0.860 & 0.0 (fixed) & 0.0 (fixed) \\
$|P_{-1}^{+}|$ & 0.130 & $-1.113$ & $-63.8$ \\
$|P_{+1}^{-}|$ & 0.142 & 0.0 (fixed) & 0.0 (fixed) \\
$|P_{-1}^{-}|$ & 0.473 & $-1.635$ & $-93.7$ \\
\hline
\end{tabular}
\end{table}

\begin{table}[h!]
\centering
\caption{Solution 2, the ambiguous partner solution.}
\begin{tabular}{|c|c|c|c|}
\hline
\textbf{Amplitude} & \textbf{Magnitude} & \textbf{Phase (radians)} & \textbf{Phase (degrees)} \\
\hline
$|P_{+1}^{+}|$ & 0.848 & 0.0 (fixed) & 0.0 (fixed) \\
$|P_{-1}^{+}|$ & 0.191 & $-1.260$ & $-72.2$ \\
$|P_{+1}^{-}|$ & 0.027 & 0.0 (fixed) & 0.0 (fixed) \\
$|P_{-1}^{-}|$ & 0.493 & $-1.890$ & $-108.3$ \\
\hline
\end{tabular}
\end{table}

\section{Relationships between $\rho$ Spin Density Matrix Elements and the Moments}
\label{App:RhoSDMEMoments}
Note on convention. In this paper we use $H^0(0,0)\equiv2$. However, for the equations in this appendix we give the using the convention $H^0(0,0)\equiv1$.
\begin{align*}
    \rho^{0}_{00} &= \frac{1}{3}(5H^{0}(20) + 1) \\
    \rho^{0}_{1-1} &= -\frac{5}{\sqrt{6}}H^{0}(22) \\
    \Re\rho^{0}_{10} &=\frac{5}{\sqrt{12}}H^{0}(21) \\[1em]
    \rho^{1}_{00} &= -\frac{1}{3}H^{1}(00) - \frac{5}{3}H^{1}(20) \\
    \rho^{1}_{1-1} &= \frac{5}{\sqrt{6}}H^{1}(22) \\
    \Re\rho^{1}_{10} &= -\frac{5}{\sqrt{12}}H^{1}(21) \\
    \rho^{1}_{11} &= -\frac{1}{3}H^{1}(00) + \frac{5}{6}H^{1}(20) \\[1em]
    \Im\rho^{2}_{1-1} &= \frac{5}{\sqrt{6}}H^{2}(22) \\
    \Im\rho^{2}_{10} &= -\frac{5}{\sqrt{12}}H^{2}(21) \\[1em]
    \Im\rho^{3}_{1-1} &= \frac{5}{\sqrt{6}}H^{3}(22) \\
    \Im\rho^{3}_{10} &= -\frac{5}{\sqrt{12}}H^{3}(21)
\end{align*}

\begin{align*}
    H^{0}(00) &= 1\\
    H^{0}(20) &= \frac{1}{5}(3\rho^{0}_{00} - 1) \\
    H^{0}(21) &= \frac{\sqrt{12}}{5}\Re\rho^{0}_{10} \\
    H^{0}(22) &= -\frac{\sqrt{6}}{5}\rho^{0}_{1-1} \\[1em]
    H^{1}(00) &= -(2\rho^{1}_{11} + \rho^{1}_{00}) \\
    H^{1}(20) &= \frac{2}{5}(\rho^{1}_{11} - \rho^{1}_{00}) \\
    H^{1}(21) &= -\frac{\sqrt{12}}{5}\Re\rho^{1}_{10} \\
    H^{1}(22) &= \frac{\sqrt{6}}{5}\rho^{1}_{1-1} \\[1em]
    H^{2}(21) &= -\frac{\sqrt{12}}{5}\Im\rho^{2}_{10} \\
    H^{2}(22) &= \frac{\sqrt{6}}{5}\Im\rho^{2}_{1-1} \\[1em]
    H^{3}(21) &= -\frac{\sqrt{12}}{5}\Im\rho^{3}_{10} \\
    H^{3}(22) &= \frac{\sqrt{6}}{5}\Im\rho^{3}_{1-1}
\end{align*}
\newpage

\section{Relating S,P Partial Waves to Moments of Spherical Harmonic Functions} \label{App:Spin1H}
This appendix provides the explicit expressions for the spherical harmonic moments $H^{\alpha}(L,M)$ in terms of S- and P-wave amplitudes up to $L=2$, with the summations over reflectivity ($\epsilon$) written out explicitly. The numerical factors are the exact analytical coefficients derived from the formalism in Ref.~[1]. The amplitudes are expressed in terms of their magnitudes ($|S|$, $|P|$) and phases ($\phi$).

\vspace{1em}
\begin{align*}
H^{0}(0,0) ={}& 2 \left( |S_{0}^{+}|^2 + |P_{+1}^{+}|^2 + |P_{0}^{+}|^2 + |P_{-1}^{+}|^2 \right) \\
& + 2 \left( |S_{0}^{-}|^2 + |P_{+1}^{-}|^2 + |P_{0}^{-}|^2 + |P_{-1}^{-}|^2 \right) \\[1em]
H^{1}(0,0) ={}& 2 \left( |S_{0}^{+}|^2 + |P_{0}^{+}|^2 - 2 |P_{+1}^{+}||P_{-1}^{+}|\cos(\phi_{P_{+1}}^{+} - \phi_{P_{-1}}^{+}) \right) \\
& - 2 \left( |S_{0}^{-}|^2 + |P_{0}^{-}|^2 - 2 |P_{+1}^{-}||P_{-1}^{-}|\cos(\phi_{P_{+1}}^{-} - \phi_{P_{-1}}^{-}) \right) \\[1em]
H^{2}(0,0) ={}& 0 \\[1em]
H^{3}(0,0) ={}& 0
\end{align*}

\hrule
\vspace{1em}

\begin{align*}
H^{0}(1,0) ={}& \frac{4\sqrt{3}}{3} \left( |S_{0}^{+}||P_{0}^{+}|\cos(\phi_{S_0}^{+} - \phi_{P_0}^{+}) + |S_{0}^{-}||P_{0}^{-}|\cos(\phi_{S_0}^{-} - \phi_{P_0}^{-}) \right) \\[1em]
H^{1}(1,0) ={}& \frac{4\sqrt{3}}{3} \left( |S_{0}^{+}||P_{0}^{+}|\cos(\phi_{S_0}^{+} - \phi_{P_0}^{+}) - |S_{0}^{-}||P_{0}^{-}|\cos(\phi_{S_0}^{-} - \phi_{P_0}^{-}) \right) \\[1em]
H^{2}(1,0) ={}& 0 \\[1em]
H^{3}(1,0) ={}& 0
\end{align*}

\hrule
\vspace{1em}

\begin{align*}
H^{0}(1,1) ={}& \frac{2\sqrt{3}}{3} \left( |S_{0}^{+}||P_{+1}^{+}|\cos(\phi_{S_0}^{+} - \phi_{P_{+1}}^{+}) - |S_{0}^{+}||P_{-1}^{+}|\cos(\phi_{S_0}^{+} - \phi_{P_{-1}}^{+}) \right) \\
& + \frac{2\sqrt{3}}{3} \left( |S_{0}^{-}||P_{+1}^{-}|\cos(\phi_{S_0}^{-} - \phi_{P_{+1}}^{-}) - |S_{0}^{-}||P_{-1}^{-}|\cos(\phi_{S_0}^{-} - \phi_{P_{-1}}^{-}) \right) \\[1em]
H^{1}(1,1) ={}& \frac{2\sqrt{3}}{3} \left( |S_{0}^{+}||P_{+1}^{+}|\cos(\phi_{S_0}^{+} - \phi_{P_{+1}}^{+}) - |S_{0}^{+}||P_{-1}^{+}|\cos(\phi_{S_0}^{+} - \phi_{P_{-1}}^{+}) \right) \\
& - \frac{2\sqrt{3}}{3} \left( |S_{0}^{-}||P_{+1}^{-}|\cos(\phi_{S_0}^{-} - \phi_{P_{+1}}^{-}) - |S_{0}^{-}||P_{-1}^{-}|\cos(\phi_{S_0}^{-} - \phi_{P_{-1}}^{-}) \right) \\[1em]
H^{2}(1,1)/i ={}& \frac{2\sqrt{3}}{3} \left( -|S_{0}^{+}||P_{+1}^{+}|\cos(\phi_{S_0}^{+} - \phi_{P_{+1}}^{+}) - |S_{0}^{+}||P_{-1}^{+}|\cos(\phi_{S_0}^{+} - \phi_{P_{-1}}^{+}) \right) \\
& - \frac{2\sqrt{3}}{3} \left( -|S_{0}^{-}||P_{+1}^{-}|\cos(\phi_{S_0}^{-} - \phi_{P_{+1}}^{-}) - |S_{0}^{-}||P_{-1}^{-}|\cos(\phi_{S_0}^{-} - \phi_{P_{-1}}^{-}) \right) \\[1em]
H^{3}(1,1)/i ={}& \frac{2\sqrt{3}}{3} \left( |S_{0}^{+}||P_{+1}^{+}|\sin(\phi_{S_0}^{+} - \phi_{P_{+1}}^{+}) + |S_{0}^{+}||P_{-1}^{+}|\sin(\phi_{S_0}^{+} - \phi_{P_{-1}}^{+}) \right) \\
& + \frac{2\sqrt{3}}{3} \left( |S_{0}^{-}||P_{+1}^{-}|\sin(\phi_{S_0}^{-} - \phi_{P_{+1}}^{-}) + |S_{0}^{-}||P_{-1}^{-}|\sin(\phi_{S_0}^{-} - \phi_{P_{-1}}^{-}) \right)
\end{align*}

\hrule
\vspace{1em}

\begin{align*}
H^{0}(2,0) ={}& 2 \left( \frac{2}{5}|P_{0}^{+}|^2 - \frac{1}{5}|P_{+1}^{+}|^2 - \frac{1}{5}|P_{-1}^{+}|^2 \right) + 2 \left( \frac{2}{5}|P_{0}^{-}|^2 - \frac{1}{5}|P_{+1}^{-}|^2 - \frac{1}{5}|P_{-1}^{-}|^2 \right) \\[1em]
H^{1}(2,0) ={}& 2 \left( \frac{2}{5}|P_{0}^{+}|^2 + \frac{2}{5}|P_{+1}^{+}||P_{-1}^{+}|\cos(\phi_{P_{+1}}^{+} - \phi_{P_{-1}}^{+}) \right) \\
& - 2 \left( \frac{2}{5}|P_{0}^{-}|^2 + \frac{2}{5}|P_{+1}^{-}||P_{-1}^{-}|\cos(\phi_{P_{+1}}^{-} - \phi_{P_{-1}}^{-}) \right) \\[1em]
H^{2}(2,0) ={}& 0 \\[1em]
H^{3}(2,0) ={}& 0
\end{align*}

\hrule
\vspace{1em}

\begin{align*}
H^{0}(2,1) ={}& \frac{2\sqrt{3}}{5} \left( |P_{+1}^{+}||P_{0}^{+}|\cos(\phi_{P_{+1}}^{+} - \phi_{P_0}^{+}) - |P_{0}^{+}||P_{-1}^{+}|\cos(\phi_{P_0}^{+} - \phi_{P_{-1}}^{+}) \right) \\
& + \frac{2\sqrt{3}}{5} \left( |P_{+1}^{-}||P_{0}^{-}|\cos(\phi_{P_{+1}}^{-} - \phi_{P_0}^{-}) - |P_{0}^{-}||P_{-1}^{-}|\cos(\phi_{P_0}^{-} - \phi_{P_{-1}}^{-}) \right) \\[1em]
H^{1}(2,1) ={}& \frac{2\sqrt{3}}{5} \left( |P_{+1}^{+}||P_{0}^{+}|\cos(\phi_{P_{+1}}^{+} - \phi_{P_0}^{+}) - |P_{0}^{+}||P_{-1}^{+}|\cos(\phi_{P_0}^{+} - \phi_{P_{-1}}^{+}) \right) \\
& - \frac{2\sqrt{3}}{5} \left( |P_{+1}^{-}||P_{0}^{-}|\cos(\phi_{P_{+1}}^{-} - \phi_{P_0}^{-}) - |P_{0}^{-}||P_{-1}^{-}|\cos(\phi_{P_0}^{-} - \phi_{P_{-1}}^{-}) \right) \\[1em]
H^{2}(2,1)/i ={}& \frac{2\sqrt{3}}{5} \left( -|P_{+1}^{+}||P_{0}^{+}|\cos(\phi_{P_{+1}}^{+} - \phi_{P_0}^{+}) - |P_{0}^{+}||P_{-1}^{+}|\cos(\phi_{P_0}^{+} - \phi_{P_{-1}}^{+}) \right) \\
& - \frac{2\sqrt{3}}{5} \left( -|P_{+1}^{-}||P_{0}^{-}|\cos(\phi_{P_{+1}}^{-} - \phi_{P_0}^{-}) - |P_{0}^{-}||P_{-1}^{-}|\cos(\phi_{P_0}^{-} - \phi_{P_{-1}}^{-}) \right) \\[1em]
H^{3}(2,1)/i ={}& \frac{2\sqrt{3}}{5} \left( |P_{+1}^{+}||P_{0}^{+}|\sin(\phi_{P_{+1}}^{+} - \phi_{P_0}^{+}) + |P_{0}^{+}||P_{-1}^{+}|\sin(\phi_{P_0}^{+} - \phi_{P_{-1}}^{+}) \right) \\
& + \frac{2\sqrt{3}}{5} \left( |P_{+1}^{-}||P_{0}^{-}|\sin(\phi_{P_{+1}}^{-} - \phi_{P_0}^{-}) + |P_{0}^{-}||P_{-1}^{-}|\sin(\phi_{P_0}^{-} - \phi_{P_{-1}}^{-}) \right)
\end{align*}

\hrule
\vspace{1em}

\begin{align*}
H^{0}(2,2) ={}& -\frac{2\sqrt{6}}{5} \left( |P_{+1}^{+}||P_{-1}^{+}|\cos(\phi_{P_{+1}}^{+} - \phi_{P_{-1}}^{+}) + |P_{+1}^{-}||P_{-1}^{-}|\cos(\phi_{P_{+1}}^{-} - \phi_{P_{-1}}^{-}) \right) \\[1em]
H^{1}(2,2) ={}& \frac{\sqrt{6}}{5} \left( |P_{+1}^{+}|^2 + |P_{-1}^{+}|^2 \right) - \frac{\sqrt{6}}{5} \left( |P_{+1}^{-}|^2 + |P_{-1}^{-}|^2 \right) \\[1em]
H^{2}(2,2)/i ={}& \frac{\sqrt{6}}{5} \left( |P_{-1}^{+}|^2 - |P_{+1}^{+}|^2 \right) - \frac{\sqrt{6}}{5} \left( |P_{-1}^{-}|^2 - |P_{+1}^{-}|^2 \right) \\[1em]
H^{3}(2,2)/i ={}& \frac{2\sqrt{6}}{5} \left( |P_{+1}^{+}||P_{-1}^{+}|\sin(\phi_{P_{+1}}^{+} - \phi_{P_{-1}}^{+}) + |P_{+1}^{-}||P_{-1}^{-}|\sin(\phi_{P_{+1}}^{-} - \phi_{P_{-1}}^{-}) \right)
\end{align*}

\section{Investigating solutions $\chi^2$} \label{App:logVal}
\begin{figure*}[tbp]
    \centering
    \includegraphics[width=\textwidth]{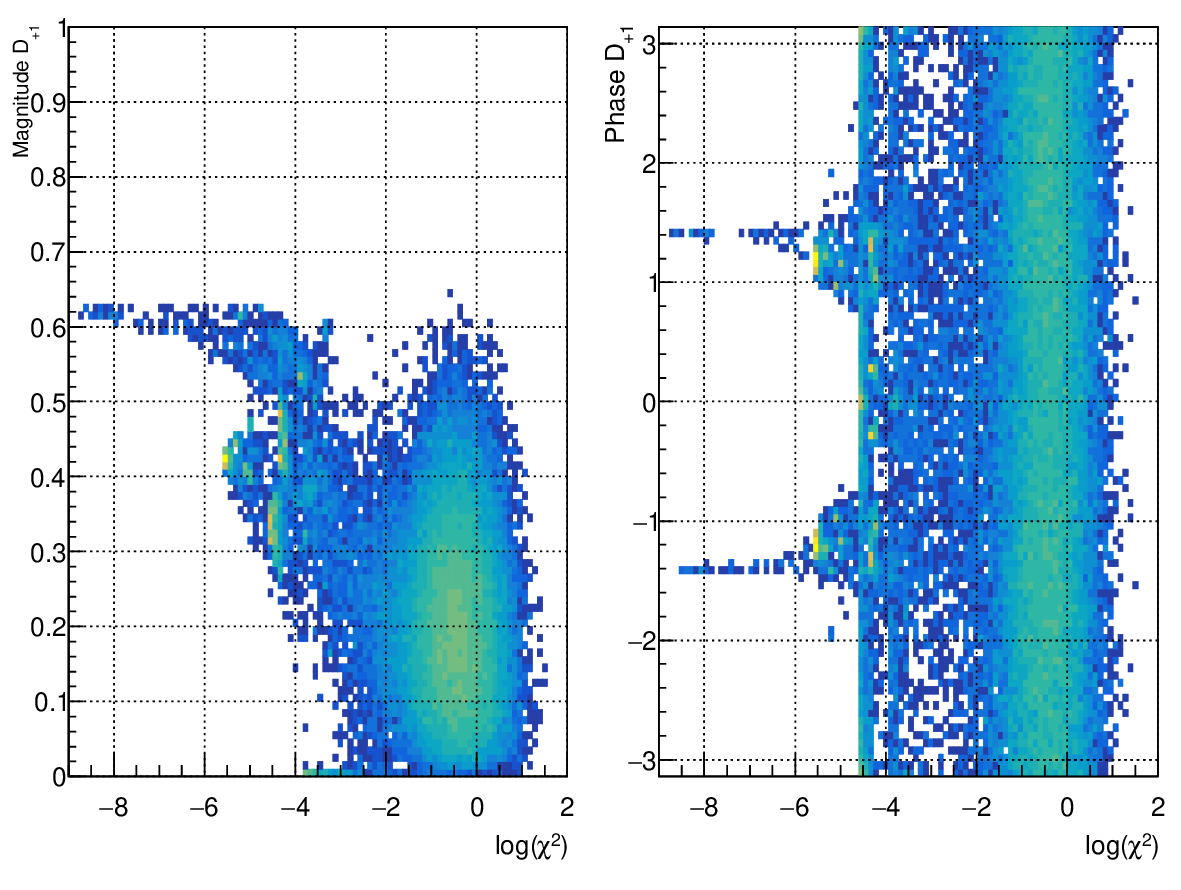}
    \caption{ The magnitude (left) and phase (right) of the $^+D_{+1}$ versus the minimisation $\log(\chi^2)$ for the numerical inversion for 50 thousand inversions. This is for the case of linearly polarized beams.
    }
    \label{fig:logval_h012}
\end{figure*}

\begin{figure*}[tbp]
    \centering
    \includegraphics[width=\textwidth]{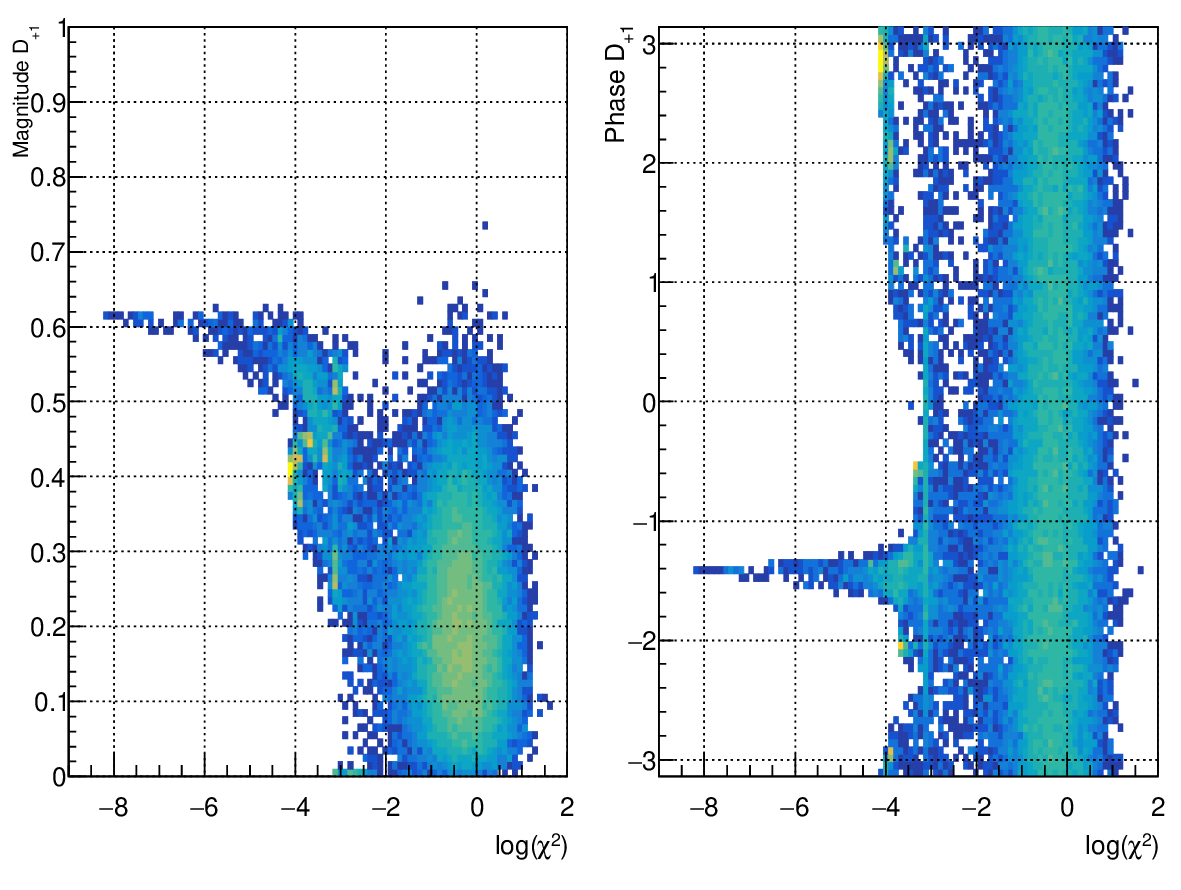}
    \caption{ The magnitude (left) and phase (right) of the $^+D_{+1}$ versus the minimisation $\log(\chi^2)$ for the numerical inversion for 50 thousand inversions. This is for the case of circularly polarized beams.      
    }
    \label{fig:logval_h0123}
\end{figure*}

In this section we examine the likelihood space, $\log(\chi^2) $, for the nominal $S$ and $D$ generated waveset, but allowing P waves and negative reflectivity waves in equation inversion. 
We choose just to show the magnitude and phase of the $D_{+1}$ wave as a function of the minimum $\chi^2$ found in each of 50,000 inversions, with each inversion having a different random starting point in amplitude space. In ~\cref{fig:logval_h012} we show for linearly polarized data, that is ${H^0,H^1,H^2}$ moments, while ~\cref{fig:logval_h0123} show the case of elliptically polarized where we also use ${H^3}$. We observe that at low values of $\log(\chi^2) < -6$ where the true solution is found we get a single pair of complex conjugate solutions for ${H^0,H^1,H^2}$  and a single solution for ${H^0,H^1,H^2,H^3}$; The ${H^0,H^1,H^2}$ distribution is symmetric in phase; there are local minima found in the region $[-6,-3]$ in the first case and $[-4,-3]$ in the second case. This shows the additional constraints from the circular polarized $H^3$ data, which therefore help distinguish true and false minima. These are still relatively low values for $\log(\chi^2)$ and would prove difficult to differentiate from the true solution in real experimental data, with various sources of noise. This was observed already in Ref.~\cite{kjcp-h8b9}. 

\section{Counting Formula for Partial Wave Truncations}
\label{App:counting}

This appendix presents the definitions and key algebraic relations used for determining the necessary degrees of freedom in the partial-wave analysis (PWA) as a function of the maximum angular momentum ($\ell_{max}$) included in the truncated model space.

\subsection{Partial Wave Amplitudes and Real Parameters}

These formulas define the number of complex partial-wave amplitudes and the corresponding number of independent real parameters required to specify the scattering matrix.

\begin{table}[h]
    \centering
    \caption{Definitions for Amplitude Parameters}
    \begin{tabular}{|c|c|c|}
    \hline
    \textbf{Quantity} & \textbf{Formula} & \textbf{Description} \\
    \hline
    $N_{waves}$ & $N_{waves} = 2(\ell_{max}+1)^2$ & Total number of complex partial waves (amplitudes) up to spin $\ell_{max}$.  \\
    \hline
    $N_{real}$ & $N_{real} = 2 \cdot N_{waves} - 2$ & Total number of independent real parameters, accounting for a fixed phase in each reflectivity. \\
    \hline
    \end{tabular}
\end{table}

\subsection{Observable Constraints (SDMEs and Moments)}

These formulas determine the total number of unique constraints provided by the scattering observables, which must be compared against $N_{real}$ to assess the system's solvability.

\begin{table}[h]
    \centering
    \caption{Definitions for Constraints (SDMEs and Moments)}
    \begin{tabular}{|c|c|p{8cm}|}
    \hline
    \textbf{Quantity} & \textbf{Formula} & \textbf{Description} \\
    \hline
    $N_{sdme}$ & $N_{sdme} = \frac{N_{waves}(N_{waves}+1)}{2}$ & Maximal number of unique bilinear combinations of partial wave amplitudes (SDMEs). \\
    \hline
    $N_{m1}$ (for $\mathcal{I}^0, \mathcal{I}^1$) & $N_{m1} = (2\ell_{max}+1)(\ell_{max}+1)$ & Number of independent, non-zero moments for the unpolarized ($\mathcal{I}^0$) and linear $\cos 2\Phi$ ($\mathcal{I}^1$) terms. \\
    \hline
    $N_{m2}$ (for $\mathcal{I}^2, \mathcal{I}^3$) & $N_{m2} = (2\ell_{max}+1)\ell_{max}$ & Number of independent, non-zero moments for the linear $\sin 2\Phi$ ($\mathcal{I}^2$) and circular ($\mathcal{I}^3$) terms. \\
    \hline
    $N_{m012}$ & $N_{m012} = 2 \cdot N_{m1} + N_{m2}$ & Total constraints from a \textbf{Fully Linearly Polarized} experiment ($\mathcal{I}^0, \mathcal{I}^1, \mathcal{I}^2$). \\
    \hline
    $N_{m0123}$ & $N_{m0123} = 2 \cdot N_{m1} + 2 \cdot N_{m2}$ & Total constraints from an \textbf{Elliptically Polarized} experiment ($\mathcal{I}^0, \mathcal{I}^1, \mathcal{I}^2, \mathcal{I}^3$). \\
    \hline
    \end{tabular}
\end{table}
\bibliographystyle{apsrev4-1}
\bibliography{refs}
\end{document}